\documentclass[iop,revtex4]{emulateapj}
\usepackage{natbib}
\usepackage{morefloats}
\usepackage{epsfig}
\usepackage{epsf} 
\usepackage{rotating}
\usepackage{graphicx}

\usepackage{epsfig}
\usepackage{subfigure}
\newcommand{\angstrom}{\mbox{\normalfont\AA} }
\newcommand {\ks} {km~s$^{-1} \;$}
\newcommand {\kss} {km~s$^{-1}$}

\newcommand{\arcm}{\ensuremath{\mathrm{^\prime}\;}}

\slugcomment{Submitted to The Astrophysical Journal Supplement Series}

\shorttitle{Dissecting the FF cluster MACS~J0416.1-2403}
\shortauthors{Balestra et al.}

\begin{document}

%% LaTeX will automatically break titles if they run longer than
%% one line. However, you may use \\ to force a line break if
%% you desire.

\title{CLASH-VLT: Dissecting the Frontier Fields galaxy cluster MACS~J0416.1-2403 \\
with $\sim800$ spectra of member galaxies}

%% Use \author, \affil, and the \and command to format
%% author and affiliation information.
%% Note that \email has replaced the old \authoremail command
%% from AASTeX v4.0. You can use \email to mark an email address
%% anywhere in the paper, not just in the front matter.
%% As in the title, use \\ to force line breaks.

\author{I.~Balestra\altaffilmark{1,2}, 
A.~Mercurio\altaffilmark{3}, 
B.~Sartoris\altaffilmark{1,4}, 
M.~Girardi\altaffilmark{1,4},
C.~Grillo\altaffilmark{5},
M.~Nonino\altaffilmark{1},
P.~Rosati\altaffilmark{6},
A.~Biviano\altaffilmark{1},
S.~Ettori\altaffilmark{7,8}, 
W.~Forman\altaffilmark{9},
C.~Jones\altaffilmark{9},
A.~Koekemoer\altaffilmark{10},
E.~Medezinski\altaffilmark{11,12},
J.~Merten\altaffilmark{13},
G.~A.~Ogrean\altaffilmark{9,14,25},
P.~Tozzi\altaffilmark{15},
K.~Umetsu\altaffilmark{16},
E.~Vanzella\altaffilmark{7},
R.~J.~van Weeren\altaffilmark{9},
A.~Zitrin\altaffilmark{11,25},
M.~Annunziatella\altaffilmark{1,4},
G.~B.~Caminha\altaffilmark{6},
T.~Broadhurst\altaffilmark{17},
D.~Coe\altaffilmark{10},
M.~Donahue\altaffilmark{18},
A.~Fritz\altaffilmark{19},
B.~Frye\altaffilmark{20},
D.~Kelson\altaffilmark{21},
M.~Lombardi\altaffilmark{22},
C.~Maier\altaffilmark{23},
M.~Meneghetti\altaffilmark{7,8},
A.~Monna\altaffilmark{2},
M.~Postman\altaffilmark{10},
M.~Scodeggio\altaffilmark{19},
S.~Seitz\altaffilmark{2,24},
B.~Ziegler\altaffilmark{23}
}

\email{balestra@oats.inaf.it}

%% Notice that each of these authors has alternate affiliations, which
%% are identified by the \altaffilmark after each name.  Specify alternate
%% affiliation information with \altaffiltext, with one command per each
%% affiliation.

\altaffiltext{1}{INAF - Osservatorio Astronomico di Trieste, via G. B. Tiepolo 11, I-34131, Trieste, Italy}
\altaffiltext{2}{University Observatory Munich, Scheinerstrasse 1, D-81679 M\"unchen, Germany}
\altaffiltext{3}{INAF - Osservatorio Astronomico di Capodimonte, Via Moiariello 16 I-80131 Napoli, Italy}
\altaffiltext{4}{Dipartimento di Fisica, Universit\`a degli Studi di Trieste, Via Tiepolo 11, I-34143 Trieste, Italy}
\altaffiltext{5}{Dark Cosmology Centre, Niels Bohr Institute, University of Copenhagen, Juliane Maries Vej 30, 2100 Copenhagen, Denmark}
\altaffiltext{6}{Dipartimento di Fisica e Scienze della Terra, Universit\`a di Ferrara, Via Saragat 1, I-44122 Ferrara, Italy}
\altaffiltext{7}{INAF - Osservatorio Astronomico di Bologna, Via Ranzani 1, I-40127 Bologna, Italy}
\altaffiltext{8}{INFN - Bologna, Via Ranzani 1, I-40127 Bologna, Italy}
\altaffiltext{9}{Harvard-Smithsonian Center for Astrophysics, 60 Garden Street, Cambridge, MA 02138, USA}
\altaffiltext{10}{Space Telescope Science Institute, 3700 San Martin Drive, Baltimore, MD 21208, USA}
\altaffiltext{11}{Cahill Center for Astronomy and Astrophysics, California Institute of Technology, MS 249-17, Pasadena, CA 91125, USA}
\altaffiltext{12}{Center for Astrophysics and Planetary Science, Racah Institute of Physics, The Hebrew University, Jerusalem 91904, Israel}
\altaffiltext{13}{Department of Physics, University of Oxford, Keble Road, Oxford OX1 3RH, U.K.}
\altaffiltext{14}{KIPAC, Stanford University, 452 Lomita Mall, Stanford, CA 94305, USA}
\altaffiltext{15}{INAF - Osservatorio Astrofisico di Arcetri, Largo E. Fermi 5, 50125 Firenze, Italy}
\altaffiltext{16}{Institute of Astronomy and Astrophysics, Academia Sinica, P. O. Box 23-141, Taipei 10617, Taiwan}
\altaffiltext{17}{Ikerbasque, Basque Foundation for Science, Alameda Urquijo, 36-5 Plaza Bizkaia, E-48011, Bilbao, Spain}
\altaffiltext{18}{Department of Physics and Astronomy, Michigan State University, East Lansing, MI 48824, USA}
\altaffiltext{19}{INAF - Istituto di Astrofisica Spaziale e Fisica cosmica (IASF) Milano, via Bassini 15, I-20133 Milano, Italy}
\altaffiltext{20}{Department of Astronomy/Steward Observatory, University of Arizona, 933 North Cherry Avenue, Tucson, AZ 85721, USA}
\altaffiltext{21}{Observatories of the Carnegie Institution of Washington, Pasadena, CA 91101, USA}
\altaffiltext{22}{Dipartimento di Fisica, Universit\`a degli Studi di Milano, via Celoria 16, I-20133 Milan, Italy}
\altaffiltext{23}{University of Vienna, Department of Astrophysics, T\"urkenschanzstr. 17, A-1180, Wien, Austria}
\altaffiltext{24}{Max-Planck-Institut f\"ur extraterrestrische Physik, Postfach 1312, Giessenbachstr., D-85741 Garching, Germany }
\altaffiltext{25}{Hubble Fellow}

\setcounter{footnote}{0}

%\date{Received 2015/ Accepted 2015}

\begin{abstract}
We present VIMOS-VLT spectroscopy of the Frontier Fields cluster 
MACS~J0416.1-2403 (z=0.397). Taken as part of the CLASH-VLT survey, the large 
spectroscopic campaign provided more than 4000 reliable redshifts
over $\sim600$ arcmin$^2$, 
including $\sim800$ cluster member galaxies. The unprecedented sample 
of cluster members at this redshift allows us to perform a highly detailed dynamical
and structural analysis of the cluster out to $\sim2.2\,r_{200}$ ($\sim 4\,$Mpc). 
Our analysis of substructures reveals a complex system composed of a main massive cluster 
($M_{200}\sim 0.9\times 10^{15}$~M$_\odot$ and $\sigma_{V,r200}\sim1000$ \kss) presenting two major 
features: i) a bimodal velocity distribution, showing two central peaks separated by 
$\Delta V_{\rm{rf}}\sim$ 1100 \ks with comparable galaxy content and velocity dispersion, ii) a 
projected elongation of the main substructures along the NE-SW direction, with a 
prominent sub-clump $\sim600$ kpc SW of the center and an isolated BCG approximately 
halfway between the center and the SW clump. We also detect a low mass structure at 
$z\sim0.390$, $\sim10'$ south of the cluster center, projected at $\sim 3$~Mpc, with a 
relative line-of-sight velocity of $\Delta V_{\rm{rf}}\sim$-1700~\ks. 
The cluster mass profile that we obtain through our dynamical analysis deviates 
significantly from the ``universal'' NFW, being best fit by a 
Softened Isothermal Sphere model instead. 
The mass profile measured from the galaxy dynamics is found to be in relatively good 
agreement with those obtained from strong and weak lensing, as well as with that from 
the X-rays, despite the clearly unrelaxed nature of the cluster. Our results reveal 
overall a complex dynamical state of this massive cluster and support the 
hypothesis that the two main subclusters are being observed in a pre-collisional phase, 
in line with recent findings from radio and deep X-ray data. With this article we also 
release the entire redshift catalog of 4386 sources in the field of this cluster, which 
includes 60 identified Chandra X-ray sources and 105 JVLA radio sources.
\end{abstract}

%% Keywords should appear after the \end{abstract} command. The uncommented
%% example has been keyed in ApJ style. See the instructions to authors
%% for the journal to which you are submitting your paper to determine
%% what keyword punctuation is appropriate.

%% Authors who wish to have the most important objects in their paper
%% linked in the electronic edition to a data center may do so in the
%% subject header.  Objects should be in the appropriate "individual"
%% headers (e.g. quasars: individual, stars: individual, etc.) with the
%% additional provision that the total number of headers, including each
%% individual object, not exceed six.  The \objectname{} macro, and its
%% alias \object{}, is used to mark each object.  The macro takes the object
%% name as its primary argument.  This name will appear in the paper
%% and serve as the link's anchor in the electronic edition if the name
%% is recognized by the data centers.  The macro also takes an optional
%% argument in parentheses in cases where the data center identification
%% differs from what is to be printed in the paper.


%% From the front matter, we move on to the body of the paper.
%% In the first two sections, notice the use of the natbib \citep
%% and \citet commands to identify citations.  The citations are
%% tied to the reference list via symbolic KEYs. The KEY corresponds
%% to the KEY in the \bibitem in the reference list below. We have
%% chosen the first three characters of the first author's name plus
%% the last two numeral of the year of publication as our KEY for
%% each reference.

%________________________________________________________________

\section{Introduction}

According to the widely accepted ``concordance model'', i.e. a cold-dark-matter dominated 
model with a cosmological constant ($\Lambda$CDM), galaxy clusters form through 
gravitational collapse of the highest density peaks in the primordial density 
fluctuations of the Universe and subsequent hierarchical assembly of smaller 
structures \citep[e.g.][]{spr06}. Being the largest structures in the Universe which have 
had time to virialise at the current epoch and retaining the different mass-energy density 
components in cosmological proportion, galaxy clusters represent important probes for 
cosmology, as well as unique natural laboratories for galaxy evolution and structure 
formation \citep[e.g.][and references therein]{dre84,kra12}.

One of the key predictions of $N$-body simulations is the self-similarity of dark-matter 
halos, which are expected to follow a ``universal'' density profile \citep[][hereafter 
NFW]{nav96,nav97}. To observationally test these theoretical predictions, good data 
quality and complementary mass probes are fundamental. First results based 
on combined strong and weak lensing mass reconstruction were controversial, finding both 
agreement with the NFW profile \citep[e.g.][]{bro05,bro052,zit09,ume10} and deviations 
from it \citep[e.g.][]{new09}. Deviations from the NFW profile seem to become apparent 
only at scales $<30$ kpc around the brightest central galaxy (BCG) and if the 
dark-matter and baryonic components can be disentangled \citep{new09,new13}. 
The combination of different, complementary probes, such as lensing,
stellar kinematic of the BCG, dynamics of cluster galaxies, and X-ray surface 
brightness, allows the reconstruction of the total mass profile of a cluster over wide 
ranges of cluster-centric distances, while keeping all the systematics inherent to 
different probes under control \citep[e.g.][]{new09,new11,biv13,von14}. 

With the aim of precisely characterizing the mass profiles of galaxy clusters using 
lensing, a sample of 25 massive clusters (20 selected as relaxed clusters based on their 
X-ray morphology and 5 selected as high-magnification lensing clusters) has been observed 
with the Hubble Space Telescope (HST) as part of the Multi-Cycle Treasury program Cluster 
Lensing And Supernova survey with Hubble \citep[CLASH; P.I.: M. Postman;][]{pos12}. 
The HST survey, nicely complemented by the Subaru wide-field imaging, has allowed 
significant improvement in reconciling theoretical predictions  on the shape of the mass 
profile and on the mass-concentration relation with observations \citep{ume14,men14,mer15}.

Targeting 14 of the southern CLASH clusters, our extensive spectroscopic campaign 
carried out with the Very Large Telescope (VLT) \citep[CLASH-VLT Large Programme 
186.A-0.798; P.I.: P. Rosati; ][]{ros14} is currently providing thousands of spectroscopic 
redshifts for cluster member galaxies and other intervening structures along the line of 
sight, including high-z, highly magnified, lensed background galaxies 
\citep[see][]{biv13,bal13,gri15,gir15}.

In the first CLASH-VLT cluster analyzed (MACS~J1206.2-0847 at $z=0.44$), the 
unprecedented number of spectroscopic redshifts of member galaxies out to the cluster 
outskirts ($\sim5\,$Mpc) has allowed to perform a mass profile reconstruction 
through galaxy dynamics well beyond the cluster virial radius \citep[see][]{biv13}. 
In that fairly relaxed cluster, the mass profile obtained from the dynamical analysis 
is found to be in extremely good agreement with that derived from all the other 
independent probes (i.e. strong lensing, weak lensing, and X-rays) over two decades 
in radius and consistent with NFW.

The combination of high-quality HST and Subaru imaging with the large VLT spectroscopic 
follow-up has allowed significant progress in many aspects of our understanding of galaxy 
cluster formation and evolution: new constraints on velocity anisotropies and 
pseudo-phase-space density profiles \citep{biv13}, contribution of mergers of different 
mass to cluster growth \citep{lem13}, pressureless of dark matter \citep{sar14}, precise 
mass profiles from weak lensing \citep{ume14}, mass-concentration relation 
\citep{men14,mer15}, substructures and galaxy population \citep{gir15}, precise stellar
mass function and stellar mass density profiles \citep{ann14}, and 
effect of the environment on galaxy evolution \citep{ann15}.

MACS~J0416.1-2403 \citep[hereafter MACS0416, ][]{ebe01} is a massive, X-ray luminous 
($L_X\sim10^{45}\,$~erg~s$^{-1}$) galaxy cluster at $z\simeq0.4$. Selected 
as one of the 5 clusters with high magnification in the CLASH sample, MACS0416
has been first imaged by the HST for 25 orbits using 16 filters as part of the CLASH 
survey \citep{pos12}, where the HST mosaics were produced following 
the approach described in \citet{koe11}. The cluster was subsequently
re-observed, as part of the Hubble Frontier Fields (HFF) initiative 
\citep{koe14,lot14}, for 140 orbits in 7 filters achieving in all of them 
a 5-sigma point-source detection limit of $\simeq$29 mag (AB).

Early works based on relatively shallow Chandra observations identified MACS0416
as a merger, possibly in a post-collisional phase, given its unrelaxed X-ray morphology and 
the observed projected separation ($\sim200\,$kpc) of the two brightest cluster galaxies 
(BCGs) \citep[see][]{man12}. 

The first strong lensing analysis performed by \citet{zit13} using CLASH HST data showed a 
quite elongated projected mass distribution in the cluster core ($\sim 250$ kpc). Combining 
weak and strong lensing analyses, \citet{jau15} detected two main central mass concentrations, 
as well as two possible secondary ones SW and NE, both at $\sim 2'$ from the cluster center. 
Comparing their mass reconstruction with the position of the X-ray peaks from the early, shallow 
Chandra observations, they measured some offset ($\sim15''$) between one of the dark matter 
concentrations and the position of the intra-cluster gas which, however, did not allow 
them to unambiguously discern between a pre-collisional or post-collisional merging scenario. 

Thanks to the large number of our CLASH-VLT spectroscopic redshifts of member 
galaxies and multiply-imaged lensed objects and to a novel technique 
to identify cluster members using multi-dimensional HST color-space information, our 
group has been able to significantly improve the strong lensing model used for mass 
reconstruction and was also able to derive a precise characterization of the sub-halo 
mass function in the core of MACS0416 \citep{gri15}. The new high-resolution mass model 
currently represents the most precise mass reconstruction available for this FF cluster. 
The model is composed of 2 cored elliptical dark-matter halos plus carefully modeled galaxy 
halos around 175 cluster members in the core. The two extended halos are 
significantly separated from the centers of the two BCGs (with projected distances of 
$\sim50$ and $\sim30$ kpc from the NE and SW BCG, respectively) and they are 
separated by a total projected distance of $\sim300$ kpc.

Recently, \citet{ogr15} compared new, deeper Chandra observations with JVLA radio data and
strong lensing mass reconstruction finding no offset between dark matter and hot baryons, as well 
as an X-ray cavity, lacking radio emission, close to the NE BCG. These findings lead to the 
conclusion that a pre-collisional merging scenario could be more likely, but did not exclude a
post-collisional configuration.

% \citep{die14} also 
%through strong lensing modeling, detect two mass concentrations, which are closer 
%together compared to the two peaks of the X-ray emission. They conclude that the 
%observed configuration can be explained by a cluster merger with a 1:1-mass-ratio 
%where the two cores are in a post-collision phase with velocity of 
%$\sim 1200$ \ks, in agreement with the relative redshift difference between the two BCGs. 

%They explain the offset between dark matter and the hot gas with projection effects 
%caused by the merger happening along the line-of-sight, but with a significant deflection 
%of the SW-component towards the NE one. 

In this paper we exploit our extensive VLT spectroscopic follow-up to investigate in detail the 
dynamical state of this complex merging cluster and to derive precise mass profile from the
cluster dynamics out to large radii, which we also compare with other mass probes available, to 
provide a very precise mass characterization over a wide range of cluster-centric distances.

The plan of the paper is the following. In Section~\ref{sam}, we describe the sample and its 
photometric and spectroscopic data. In Section~\ref{memb}, we describe the selection of 
cluster members and in Section~\ref{struct} the structural analysis of the cluster.
The results of our dynamical analysis and the mass profile derived from it are 
presented in Section~\ref{s:mass}. In Section~\ref{dis} we discuss our results and 
in Section~\ref{con} we summarize our conclusions.

Throughout the paper, magnitudes are given in the AB system (AB~$\equiv 31.4 -
2.5\log\langle f_\nu / \mathrm{nJy} \rangle$) and errors at the 68\% confidence
level, unless otherwise stated. We assume a cosmology with $\Omega_{\rm tot}, 
\Omega_M, \Omega_\Lambda = 1.0, 0.3, 0.7$ and $H_0 = 70$~km~s$^{-1}$~Mpc$^{-1}$. 
In the adopted cosmology, 1\arcm corresponds to $\sim 321$ kpc at the cluster redshift.

%__________________________________________________________________

\section{The data sample}\label{sam}

\begin{table}
\caption{Log of VIMOS observations of the Frontier Fields cluster 
MACS~J0416.1-2403, taken as part of our CLASH-VLT spectroscopic campaign.}
\label{logobs}
\begin{center}
\begin{tabular}{l c c}
\hline
\hline
Mask ID               & Date      & Exp. Time (s) \\
(1)                   &  (2)      & (3)  \\
\hline
Low-resolution masks  &               &            \\
MOS\_M0416\_LRB\_1\_M1r    & Dec 2012  & $3\times1200$ \\
MOS\_M0416\_LRB\_3\_M1     & Dec 2012  & $3\times1200$ \\
MOS\_M0416\_LRB\_2\_M1r    & Jan 2013  & $3\times1200$ \\
MOS\_M0416\_LRB\_1\_M2     & Feb 2013  & $2\times900$ \\
MOS\_M0416\_LRB\_4\_M2     & Feb 2013  & $2\times900$ \\
MOS\_M0416\_LRB\_4\_M1     & Feb 2013  & $3\times1200$ \\
MOS\_M0416\_LRB\_3\_M2     & Feb 2013  & $3\times1200$ \\
MOS\_M0416\_LRB\_2\_M2     & Feb 2013  & $3\times1200$ \\
MOS\_M0416\_LRB\_2\_M2\_2  & Sep 2013  & $3\times1200$ \\
MOS\_M0416\_LRB\_1\_M3     & Oct 2013  & $3\times1200$ \\
MOS\_M0416\_LRB\_2\_M3     & Oct 2013  & $3\times1200$ \\
MOS\_M0416\_LRB\_3\_M2\_2  & Nov 2013  & $3\times1200$ \\
MOS\_M0416\_LRB\_3\_M3     & Jan 2014  & $3\times1200$ \\
MOS\_M0416\_LRB\_4\_M1\_2  & Jan 2014  & $3\times1200$ \\
MOS\_M0416\_LRB\_4\_M3     & Jan 2014  & $3\times1200$ \\
\hline
Medium-resolution masks    &           &            \\
MOS\_M0416\_MR\_1\_M3      & Jan 2014  & $3\times1200$ \\
MOS\_M0416\_MR\_2\_M3      & Jan 2014  & $3\times1200$ \\
MOS\_M0416\_MR\_4\_M3      & Feb 2014  & $3\times1200$ \\
MOS\_M0416\_MR\_3\_M3      & Feb 2014  & $3\times1200$ \\
MOS\_M0416\_MR\_1\_M4      & Sep 2014  & $3\times800$ \\
MOS\_M0416\_MR\_4\_M4      & Nov 2014  & $3\times1200$ \\
\hline
\end{tabular}
\\
\end{center}
\textbf{Notes.} Columns list the following information: (1) mask identification number, 
(2) date of the observations, and (3) number of exposures and integration time of single
exposures.
\end{table}

%__________________________________________________________________

\subsection{CLASH-VLT spectroscopy with VIMOS}\label{specsam}

The cluster MACS~J0416.1-2403 was observed between December 2012 and November 2014 
as part of the ESO Large Programme 186.A-0798 ``Dark Matter Mass Distributions of 
Hubble Treasury Clusters and the Foundations of $\Lambda$CDM Structure Formation Models'' 
(P.I.: Piero Rosati) using VIMOS \citep{lef03} at the ESO VLT. The log of our CLASH-VLT 
observations of this cluster is in Table~\ref{logobs}. The VIMOS observations
were designed in sets of four separate pointings each with a different quadrant centered 
on the cluster core. The overlapping quadrants on the cluster core were used both to achieve 
longer integration times on faint arcs and other interesting strong lensing features and to 
have the largest possible number of slits on candidate cluster members in the crowded 
central region of the cluster.  

\begin{figure}
\centering
\includegraphics[width= 8.8cm]{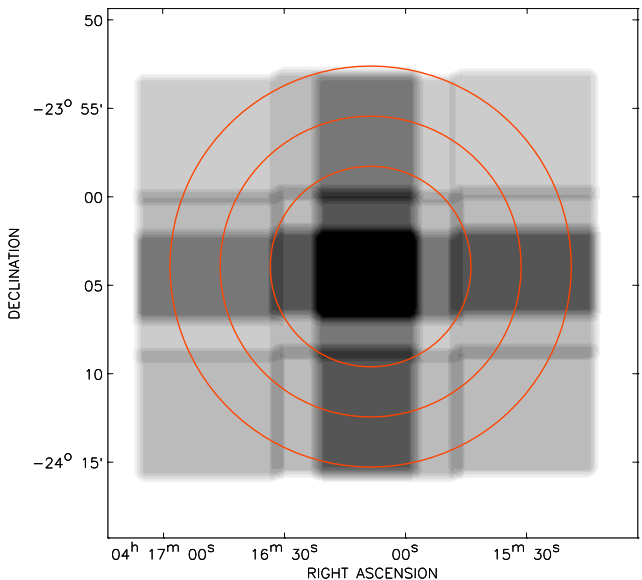}
\caption{Footprint of our CLASH-VLT spectroscopic observations shown as the resulting 
``exposure map'' of the 21 VIMOS pointings in the field of MACS0416. The size of the 
footprint is $26'\times23'$. The three red circles are centered on the 
NE BCG and have radii of 1, 1.5, and 2 $r_{200}$ (where $r_{200}=1.82$ Mpc).}
\label{expm}
\end{figure}

A total of 21 masks were observed (15 LR-Blue masks and 6 MR masks). Each mask has an 
integration time of 1 hour (split into $3\times20$ minute OBs), with the exception of 
3 masks, which have integration times of 30 and 40 minutes (see Table~\ref{logobs}). 
The footprint of our CLASH-VLT spectroscopic observations is displayed in Figure~\ref{expm}
which also shows the resulting ``exposure map'' of the 21 VIMOS pointings.

The LR-Blue masks cover the spectral range 3700-6700\angstrom with a resolution R=180, 
while the MR masks cover the range 4800-10000\angstrom with a resolution R=580.
The width of the slits is set to $1''$. To maximize the number of targets per pointing, both in 
the LR-Blue and the MR masks, we used shorter than standard slits (down to $6''$). 
This has the advantage of  improving multiplexing (we are able to fit up to 140 slits 
in a single quadrant, i.e. more than 500 targets per mask) without compromising sky 
subtraction \citep[see][]{sco09}. 

Data reduction has been performed using the VIMOS Interactive Pipeline Graphical Interface 
\citep[VIPGI, ][]{sco05}, which follows standard MOS data reduction steps (e.g. bias 
subtraction, flat-field correction, bad-pixel cleaning, sky subtraction, fringing correction, 
wavelength calibration).

Redshift determination was performed following a two-step procedure similar to the 
one used in \citet{bal10}: 
\begin{enumerate}
\item we first ran the $EZ$ software \citep{gar10} for automatic cross-correlation with 
template spectra (i.e. ordinary S0, Sa, Sb, Sc, and elliptical galaxies at low redshift, 
Lyman break galaxies, and quasars at high redshift); 
\item we visually inspected redshift solutions obtained in the first step and we modified the most 
obvious failures, where more than 2 spectral features could be unambiguously identified.   
\end{enumerate}
During the visual check we also assign a Quality Flag (QF) to each redshift, which qualitatively 
indicates the reliability of a redshift measurement. We define four redshift quality classes 
using the following criteria:
\begin{itemize}
\item ``Secure'' (QF=3), several emission lines and/or strong absorption features are 
clearly identified;
\item ``Likely''(QF=2), intermediate quality spectra where at least two spectral features 
are well identified, for instance one emission line plus at least one absorption feature;
\item ``Insecure'' (QF=1), low S/N spectra, i.e. spectral features, either in emission or 
in absorption, are less clearly identified;
\item ``Emission-line'' (QF=9), redshift based on a single emission line only.
\end{itemize}
 
To assess the reliability of these four quality classes we 
compared pairs of duplicate observations having at least one secure measurement. 
In this way we could quantify the reliability of each quality class as follows: redshifts 
with QF=3 are correct with a probability of $>99.99$\%, QF=9 with $\sim92$\% probability, 
QF=2 with $\sim75$\% probability, and QF=1 with $<40$\% probability. 
In this paper we will only consider redshifts with QF=3, 2, and 9.

\begin{figure}
\centering
\includegraphics[width= 8.8cm]{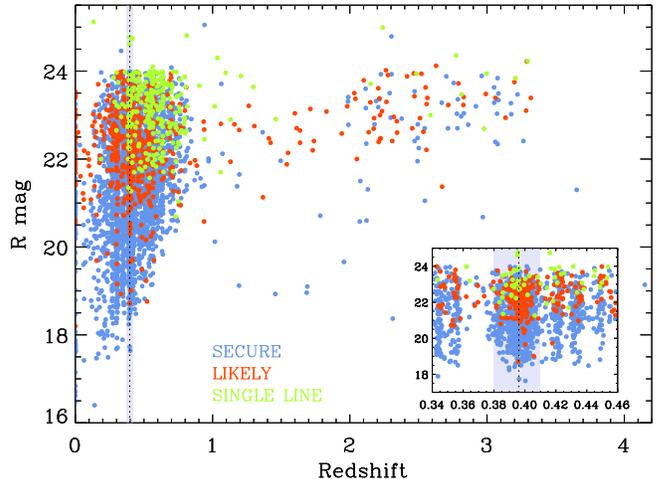}
\caption{Subaru R-band magnitude versus Redshift for all the galaxies with
reliable spectroscopic redshift. Data points are plotted in different colors according to 
the reliability classes of the spec-z, as defined in the text. The vertical dotted line 
marks the mean redshift of galaxies in MACS0416. The gray shaded area indicates
the redshift interval corresponding to rest-frame l.o.s. velocities of $\pm 3000$ km~s$^{-1}$. 
The inset is a zoom around the redshift of MACS0416.}
\label{magz}
\end{figure}

The final spectroscopic data set contains a total of 4386 reliable redshifts within an 
area of $26\times23$ arcmin around the center of the cluster, covering almost entirely the 
Subaru field of view. In Figure~\ref{magz} we plot the Subaru R-band magnitude as 
a function of the spectroscopic redshift for the entire CLASH-VLT sample of objects in 
the field of MACS0416.

\subsection{Target selection}

\begin{figure}
\centering
\includegraphics[width= 8.8cm]{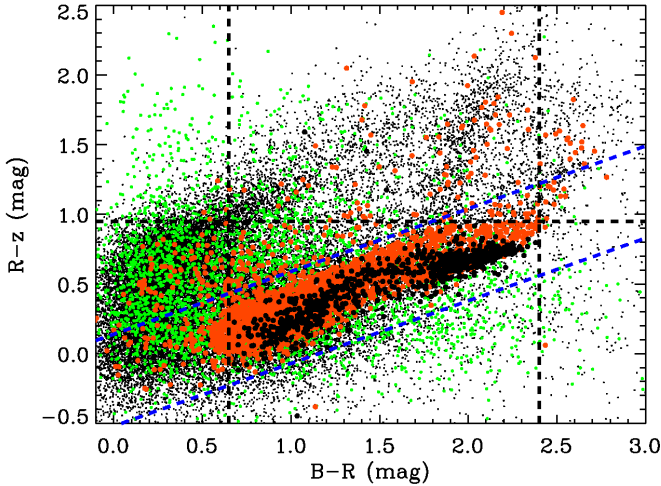}
\caption{Subaru Rc$-$z vs. B$-$Rc colors for all the sources with Rc magnitudes $<24$. 
The dashed lines show the color cuts defining the box used for target selection.
Small black dots refer to the Subaru photometric catalog. Larger colored data points mark 
objects belonging to our CLASH-VLT spectroscopic catalog (stars in green, cluster member galaxies 
in black, all the rest in red).}
\label{colcol}
\end{figure}

Targets were selected through specifically defined cuts in the color-color space using 
Subaru photometry. Figure~\ref{colcol} shows the selection box in the Rc$-$z vs. B$-$Rc 
(Subaru bands) diagram, which we adopted for the search of cluster members. The adopted 
cuts are the following:
\begin{itemize}
\item $-0.52+0.45\,(\mathrm{B}-\mathrm{Rc}) < 
\mathrm{Rc}-\mathrm{z} < 0.14+0.45\,(\mathrm{B}-\mathrm{Rc}) $ 
\item $0.65 < \mathrm{B}-\mathrm{Rc} < 2.40 $ 
\item $\mathrm{Rc}-\mathrm{z} < 0.95 $ 
\item $18.0 < \mathrm{Rc} < 24.0$
\end{itemize}
These color cuts have been chosen 
in order to maximize the success rate in cluster member selection, keeping at the same 
time the number of contaminants low and allowing for the inclusion of blue, star-forming 
cluster members in the spectroscopic sample. Notice that a fraction of objects with 
spectroscopic redshifts fall outside the selection box in Figure~\ref{colcol}. These are
mostly sources that have been manually inserted in slits, serendipitous objects found
in some slits, and a few objects with redshifts from the literature 
\citep[][ D.~Kelson, private comm.]{ebe14}. 
In addition to cluster members, we also targeted multiply-imaged lensed 
galaxies, or other strong lensing features, high-z candidates 
\citep[e.g., dropout from ][]{bra13} with $z_{phot}\lesssim 7$, a few possible 
supernova hosts \citep[e.g., ][]{pat14}, and all the Chandra sources in the field. 
Obviously, all these sources may also lie outside the selection box. 

\subsection{Catalog of spectroscopic redshifts} 

In our released CLASH-VLT spectroscopic redshift catalog (see Table~\ref{zcat}), 
for each object we provide:
(Col.~1) coordinate-based CLASH-VLT identification number, (Cols.~2, 3) coordinates 
(Subaru Suprime-Cam WCS), (Col.~4) spectroscopic redshift, (Col.~5) redshift Quality Flag, 
as described in Section~\ref{specsam}, (Col.~6) reference for the redshift (i.e.  
CLASH-VLT VIMOS [4370 redshifts]: 1=based on LR-Blue spectra [3582], 2=based on MR 
spectra [486], 3=based on a combination of LR-Blue and MR spectra [302],
4=\citet{ebe14} [5], 5=Magellan, (Kelson, private comm.) [11]), and (7) Kron R-band 
magnitude (AB, Subaru Suprime-Cam). The spectroscopic redshift 
catalog is made publicly available to the scientific community\footnote{The full CLASH-VLT 
spectroscopic catalog is publicly available on the electronic edition of this Journal, 
on CDS, and at the following URL: 
\textit{https://sites.google.com/site/vltclashpublic/}}. For repeated observations of the same
objects the catalog provides a mean value of the redshift measurements. Uncertainties on the 
redshifts vary between 75 and 150 km~s$^{-1}$, depending on the spectral resolution and
on the number and resolution of spectra on which the mean redshift is computed 
\citep[see also][]{biv13}. This catalog includes also redshifts of arcs and lensed 
multiply-imaged sources used in the strong lensing mass reconstruction presented 
by \citet{gri15}. The total number of objects in our released catalog is 4386, including 
5 redshifts from the literature \citep{ebe14} and 11 unpublished redshifts from Magellan 
observations (D.~Kelson, private comm.). The number of VIMOS spectra obtained with 
the MR grism amounts to $\sim15\%$ of the total. In the released catalog, where a 
single entry is preserved in case of duplicate observations of the same object, 3582 
redshifts are measured from LR-Blue spectra, 486 from MR spectra, and 302 from a 
combination of LR-Blue and MR spectra.

\begin{table*}
\caption{CLASH-VLT spectroscopic redshift catalog of MACS~J0416.1-2403.}
\label{zcat}
\begin{center}
\begin{tabular}{l c c c c c c c}
\hline
\hline
ID      & RA    & Dec & $z$ & QF & Ref. & Mag  \\
(1)     &  (2)    & (3)  & (4)   & (5) & (6) & (7)  \\
\hline
\hline
CLASHVLTJ041531.2-241516  &  63.879843  &  -24.254507  &  0.4648  & 3  & 1 & 19.32 \\
CLASHVLTJ041615.9-241539  &  64.066112  &  -24.261057  &  0.3448  & 3  & 1 & 19.92 \\
CLASHVLTJ041614.3-241542  &  64.059571  &  -24.261808  &  0.3021  & 3  & 1 & 20.93 \\
CLASHVLTJ041654.3-241545  &  64.226374  &  -24.262606  &  0.3078  & 3  & 1 & 21.94 \\
CLASHVLTJ041521.3-241510  &  63.838858  &  -24.252900  &  0.3009  & 3  & 1 & 19.60 \\
CLASHVLTJ041610.1-241543  &  64.041928  &  -24.262141  &  0.7231  & 3  & 1 & 21.95 \\
CLASHVLTJ041543.4-241517  &  63.930875  &  -24.254946  &  0.1902  & 3  & 1 & 19.56 \\
CLASHVLTJ041614.1-241547  &  64.058570  &  -24.263075  &  0.3009  & 3  & 1 & 22.61 \\
CLASHVLTJ041632.5-241547  &  64.135563  &  -24.263133  &  0.5082  & 3  & 1 & 23.06 \\
CLASHVLTJ041619.0-241545  &  64.079154  &  -24.262658  &  0.2999  & 2  & 1 & 22.37 \\
\hline
\end{tabular}
\end{center}
\begin{footnotesize}\textbf{Notes.} Only a portion of the table is shown here to 
illustrate the form and content of the catalog. The entire table is available in the 
electronic edition of this Journal, on CDS, and at following URL: 
\textit{https://sites.google.com/site/vltclashpublic/}. The full table 
contains 7 columns and 4386 redshifts. Columns list the following information: (1)
VIMOS identification number, (2-3) coordinates, , (4) spectroscopic redshift, (5) 
redshift quality flag, (6) reference (i.e. CLASH-VLT VIMOS [4370 redshifts]: 
1=based on LR-Blue spectra [3582], 2=based on MR spectra [486], 3=based on a combination 
of LR-Blue and MR spectra [302], 4=\citet{ebe14} [5], 5=Magellan (Kelson, private comm.) 
[11]), and (7) Subaru R-band magnitude.
\end{footnotesize}
\end{table*}

\subsection{Completeness of the spectroscopic catalog}

\begin{figure}
\centering
\includegraphics[width= 8.8cm]{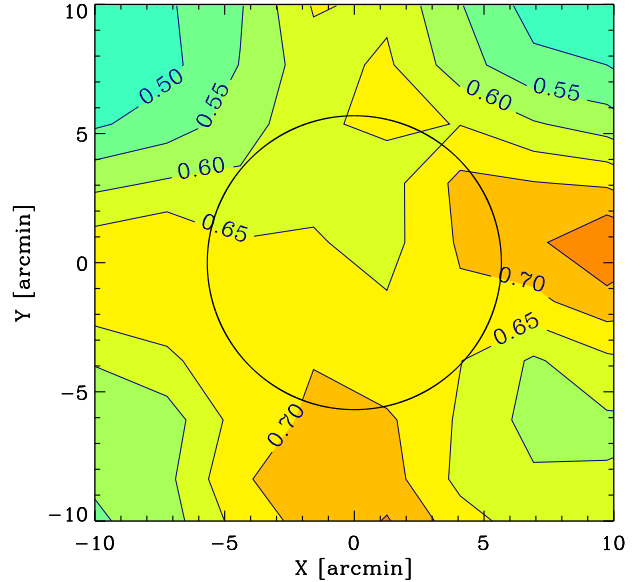}
\caption{Completeness map of our CLASH-VLT spectroscopic sample. The contour 
levels are labeled with the value of completeness. Overall, the completeness is relatively 
uniform being mostly between 0.5 and 0.7 and declining only at the corners of the field 
observed, beyond $\sim2r_{200}$. The black circle shows the size of the cluster virial 
radius, $r_{200}=1.82$~Mpc.}
\label{complm}
\end{figure}

Inhomogeneity and incompleteness of a spectroscopic sample may affect the 
ability of detecting substructures in galaxy clusters, as well as the determination of 
the projected number density of cluster galaxies that is used to reconstruct the cluster 
mass profile from the kinematics. We checked the completeness of our spectroscopic catalog 
both as a function of position on the sky and magnitude. Figure~\ref{complm} shows the 
completeness map of our spectroscopic catalog, which we computed as the ratio of the 
2D density of objects for which we were able to obtain a redshift over the 2D density of 
the targeted objects. 
This plot shows how the completeness of our spectroscopic sample is relatively uniform 
across the field observed, declining only at the corners and beyond 
$\sim2r_{200}$. As it can be inferred from this map, the completeness is approximately 
constant as a function of radius out to $\sim2r_{200}$. The dependence of the 
completeness on magnitude follows very closely the one derived for the first CLASH-VLT 
cluster analysed \citep[see Figure 4 in][]{biv13}.

%__________________________________________________________________
\section{Selection of cluster members} \label{memb}

We will now describe the procedure used for the identification of cluster member 
galaxies in MACS0416.
Cluster members were selected among the 4386 galaxies with reliable
spectroscopic redshifts, by applying the two-steps method ``peak+gap''
(P+G) also used for the dynamical analysis of the CLASH cluster
MACS~J1206.2-0847 in \citet[][]{biv13} and in \citet{gir15}. 
This method is a combination of the 1D-DEDICA
\citep{pis93,pis96}, which is an adaptive kernel method for the
evaluation of the density probability function underlying an
observational discrete data set, and the ``shifting gapper'' method,
which uses both position and velocity information \citep{fad96,gir96}.

\begin{figure}
\centering
\includegraphics[width= 8.2cm]{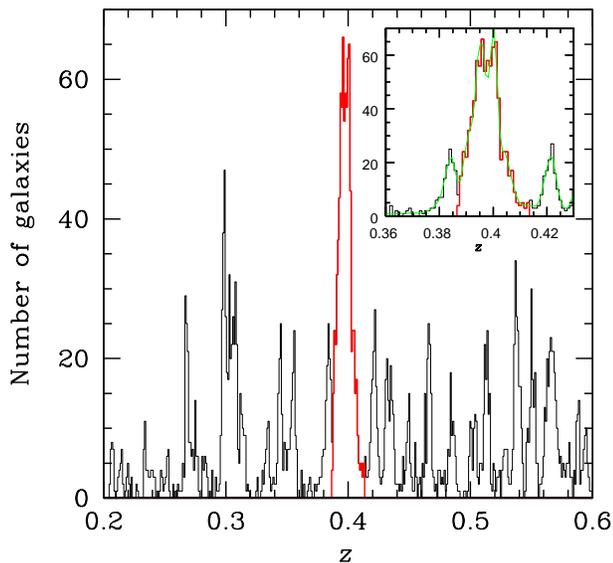}
\caption{Redshift distribution of the whole sample of galaxies
  with spectroscopic redshift $0.2<z<0.6$ (black histogram). The
  829 galaxies assigned to MACS0416 according to the DEDICA reconstruction 
  method are shown in red. The inset shows a zoom in the distribution 
  around the peak with the DEDICA reconstruction superimposed (green).
}
\label{fighisto}
\end{figure}

In the redshift distribution of our spectroscopic data of MACS0416 the
1D-DEDICA procedure detects two, largely overlapped peaks at
$z\sim0.396$ and $z\sim0.400$, composed of 428 and 401 galaxies,
respectively (see Figure~\ref{fighisto}). 

The second step of our member-selection procedure consists in
rejecting galaxies with line-of-sight velocities that are too far off the main body of
galaxies ($V=cz$). This is done by considering a fixed bin that shifts
along the distance from the cluster center. The procedure is iterated
until the number of cluster members converges. We
used a velocity gap of $500$ \ks in the cluster rest-frame and a bin
of 0.4~Mpc, or large enough to include at least 15 galaxies. Notice
that the adopted parameters here are much more stringent than those
originally used in \citet{fad96}, owing to the much better sampling
and much higher density of galaxies in our spectroscopic sample.

\begin{figure}
\centering
\includegraphics[width=8.8 cm]{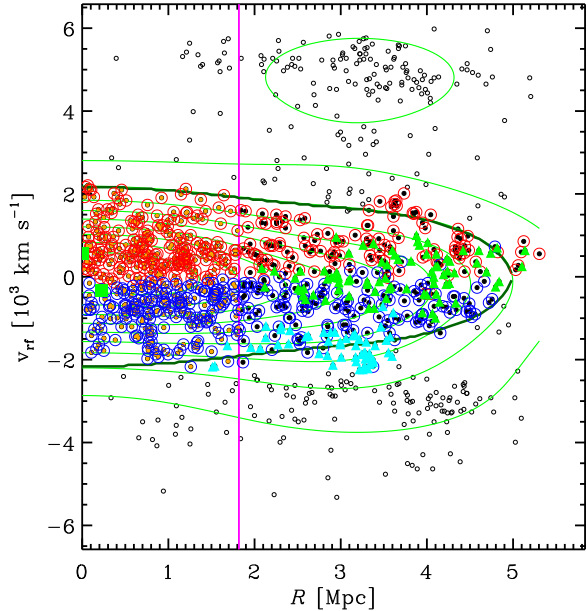}
\caption{Projected phase-space diagram: rest frame velocities are plotted as 
a function of the projected radii. Filled dots identify galaxies selected as cluster members. 
Blue and red circles mark galaxies in the two main subclusters.
Cyan triangles indicate fiducial members belonging to the Sext group, while 
green triangles are for galaxies of the less prominent W substructure detected 
in the 3D analysis described in Section~\ref{struct}. The two green squares mark 
the positions of the two BCGs. The vertical, magenta line indicates the value of
$r_{200}=1.82$~Mpc, obtained from the weak lensing analysis by \citet{ume14}. 
The green curves show the isodensity contours and the thick, dark-green line 
is the Caustic selected for our dynamical analysis (see Section~\ref{s:mass}).}
\label{figvd}
\end{figure}

Given the complex structure of the core, with two BCGs and two extended dark matter halos
separated by $\sim40-50''$ and aligned along the NE-SW direction, defining the center 
of the cluster is not obvious. As shown in \citet{gri15}, the barycenter of the cluster lies 
approximately halfway along the line connecting the two BCGs 
(R.A.=04:16:08.60, Dec.=-24:04:25.2), $\sim 23''$ from NE-BCG. 
However, the X-ray emission clearly shows a more prominent peak coincident with the 
position of the NE-BCG \citep[see][]{ogr15}. We decided to adopt the position 
of the NE-BCG and the peak of the X-ray emission
(R.A.=04:16:09.14, Dec.=-24:04:03.1) as the center of the cluster throughout our analysis. 
With this choice of the center, the second step of our member-selection procedure 
rejects 48 galaxies. The remaining 781 galaxies represent our fiducial sample of 
cluster members. Figure~\ref{figvd} shows the projected phase-space diagram (i.e. rest
frame line-of-sight velocities vs. projected radii) and the isodensity contours 
(i.e. ``Caustics''). 

We also verified that, for different choices of the center, galaxies identified as 
cluster members do not vary significantly. For example, 
if the center is fixed on the position of the barycenter \citep{gri15}, 
the galaxies identified as members are identical to those of our fiducial sample, 
except for about 10 galaxies ($\sim1\%$), which are all lying at radii larger than $5'$.
Therefore, the exact choice of the center has little influence on the results of our 
dynamical analysis. Figure~\ref{figvd} also highlights the large velocity difference between 
the two BCGs ($\Delta V_{\rm{rf}}\sim 900$ \ks in the velocity rest-frame).

By applying the biweight estimator \citep{bee90} to the 781 cluster members of the whole
system, we obtain a mean cluster redshift of $\left<z\right>=0.3972\pm0.0001$. 
We estimate the line-of-sight velocity 
dispersion, $\sigma_{V}$, by using the biweight estimator and by applying the 
cosmological correction and the standard correction for velocity
errors \citep{dan80}. We obtain $\sigma_V=996_{-36}^{+12}$ \kss, where
errors are estimated through a bootstrap technique. Table~\ref{tabv}
lists the kinematical properties of the whole sample of cluster
members, as well as those of the main system and the prominent
southern external substructure as discussed in the following
Section.

\begin{table}
\caption{Kinematical properties of the whole cluster sample, the main sample (MS),
and the southern external structure (Sext).}
\label{tabv}
\begin{center}
\begin{tabular}{l r l l}
\hline
\hline
Sample          & N$_g$ & $<V>$           & $\sigma_{V}$ \\
                &       &  [km s$^{-1}$]  & [km s$^{-1}$] \\
 (1)            &  (2)  &  (3)            & (4)  \\
\hline
Whole Sample    & 781   & $119083\pm36$   & $996_{-36}^{+12}$\\
MS$^a$          & 728   & $119233\pm34$   & $919_{-30}^{+16}$\\
Sext            & 53    & $116869\pm47$   & $336_{-47}^{+17}$\\
\hline
\end{tabular}
\end{center}
\begin{footnotesize}\textbf{Notes.} Columns list the following information: (1) id of 
the subsample, (2) number of galaxies, (3) average velocity, and 
(4) velocity dispersion. \\
$^a$ When restricting the analysis to galaxies of the MS within $r_{200}$, we measure 
$\sigma_V=998^{+25}_{-39}$~\kss.
\end{footnotesize}
\end{table}

%__________________________________________________________________
\section{Structural analysis} \label{struct}

In this Section we investigate the complex structure of the cluster, using the large 
spectroscopic sample of cluster members to determine the presence of substructures 
within the cluster.

\begin{figure}
\centering
\includegraphics[width=8.0cm]{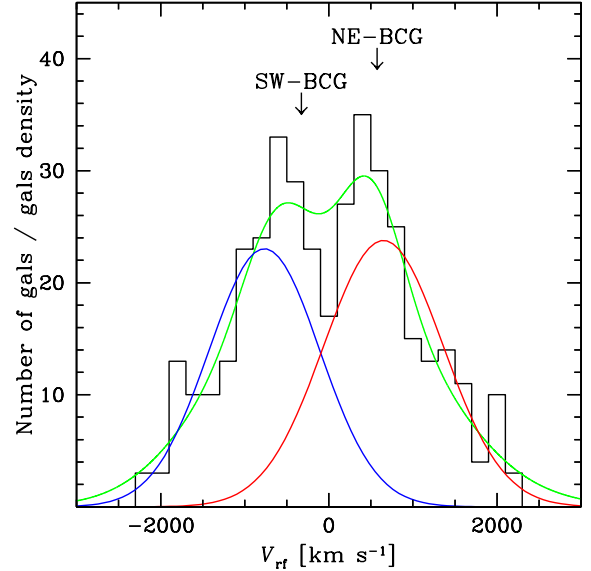}
\caption{Rest-frame velocity distribution and 1D-DEDICA density reconstruction for all
the galaxies within $r_{200}$ (black histogram and green curve). The two arrows 
indicate the velocities of the two BCGs. The two Gaussians obtained by applying 
the KMM method are also displayed (blue and red curves).}
\label{fighistocl}
\end{figure}

The first, clearly apparent feature of the cluster structure is a bimodality of the 
velocity distribution of cluster members. The bimodality is evident both in the whole 
galaxy sample and in the subsample of cluster members within the virial radius. 
This is illustrated in Figure~\ref{fighistocl}, which shows a histogram of the velocity
distribution of the 388 cluster galaxies lying within $r_{200}$
together with the 1D-DEDICA reconstruction. The 1D-DEDICA method
assigns 181 and 207 members to the low- and high-velocity subcluster,
respectively. The two subclusters, which are largely overlapping (having 238 out
of the 388 galaxies in common), peak at $z=0.395$ and $z=0.400$,
respectively. Their redshift difference corresponds to $\Delta V_{\rm
rf}\sim 1100$~\kss. Table~\ref{tabv} lists the kinematical properties 
of the whole sample of cluster members, as well as those of the main system (MS) 
and the prominent southern external substructure (Sext), separately.  

We also applied the Kaye's mixture model (KMM) test \citep{ash94,gir08}, which favors a 
two-group partition over a single Gaussian, extracting two groups of comparable galaxy 
content and velocity dispersion ($\sim700$ \kss) at the $\sim95\%$ confidence
level. Both BCGs have a significant peculiar velocity ($> 99\%$ c.l.)
with respect to the velocity distribution of the galaxy sample within
$r_{200}$ according to the indicator test by \citet{geb91}.

\begin{figure}
\centering
\includegraphics[width=8.0cm]{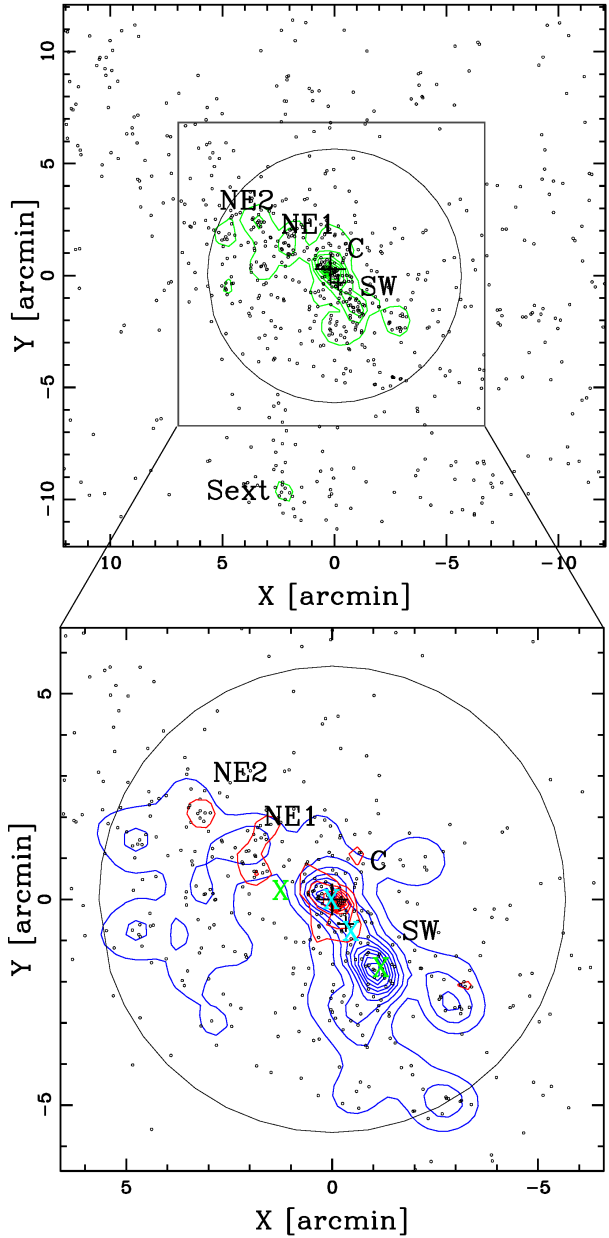}
\caption{{\em Upper panel:} spatial distribution on the sky and relative isodensity 
contour map of the 781 spectroscopic members of MACS0416
(Whole Sample), obtained with the 2D-DEDICA method. The positions of
the two BCGs (NE-BCG and SW-BCG) are indicated with two black
crosses. The labels mark the five main density peaks detected by 2D-DEDICA (see 
also Table~\ref{tabdedica2d}).
{\em Lower panel:} a zoomed-in version of the same plot within the virial region, where 
blue and red contours mark the isodensity of galaxies in the low- and
high-velocity subcluster, respectively. Green `X' symbols indicate the positions of 
the two mass concentrations detected though gravitational lensing by \citet{jau15}
(S1 at SW and S2 at NE) and cyan `X' symbols mark the position of the two peaks in
X-ray emission following \citet{ogr15}. For visual reference, we also plot a 
circle of radius $r_{200}=1.82$~Mpc, as obtained from the weak lensing analysis 
by \citet{ume14}.}
\label{figk2g}
\end{figure}

We used the 2D-DEDICA method to detect density peaks in projection on
the plane of the sky. In Figure~\ref{figk2g} (upper panel) we plot the spatial 
distribution on the sky and isodensity contours of the 781 spectroscopic members of 
MACS0416. This plot highlights the complex 2D-structure of MACS0416
and the presence of a density peak in the Sext substructure. 
Table~\ref{tabdedica2d} lists the results of
the 2D-DEDICA analysis for the four peaks with relative density
$\rho_{\rm S}>0.2$ within $r_{200}$ (C, SW, NE1, NE2), and the
densest peak outside $r_{200}$ (Sext).

\begin{table}
\caption{2D-DEDICA results on the detection of spatial substructure in the whole sample
of cluster members.}
\label{tabdedica2d}
\begin{center}
\begin{tabular}{l r c c r}
\hline
\hline
Sub-clump  & N$_{ S}$ & $\alpha$(J2000) $\delta$(J2000) & $\rho_S$ & $\chi^2_S$ \\
 (1)      &  (2)     & (3)                             & (4)      & (5)  \\
\hline
C        & 104 & 04:16:08.6 -24:04:07  & 1.00 & 55 \\
SW       & 47  & 04:16:04.2 -24:05:46  & 0.42 & 25 \\
NE1      & 44  & 04:16:17.7 -24:02:59  & 0.28 & 21 \\
NE2      & 25  & 04:16:23.3 -24:01:57  & 0.24 & 13 \\
Sext     & 29  & 04:16:18.2 -24:14:01  & 0.13 & 12 \\
\hline
\end{tabular}
\end{center}
\begin{footnotesize}\textbf{Notes.} Columns list the following information: 
(1) id of sub-clump, (2) number of assigned members, (3) coordinates of the density peak, 
(4) density relative to the densest peak, and (5) $\chi^2$ value for 
each peak.
\end{footnotesize}
\end{table}

\begin{figure}
\centering 
\includegraphics[width=9.0cm]{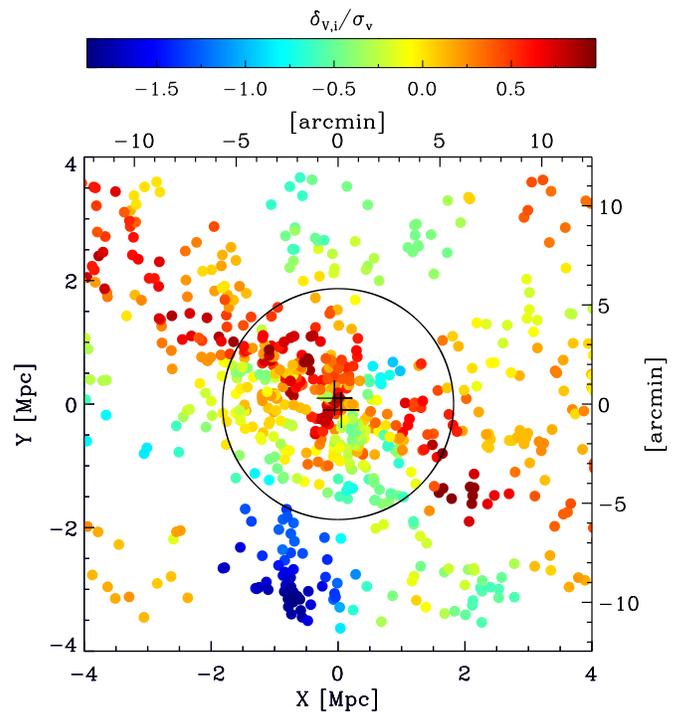}
\caption{Spatial distribution of the 781 cluster-member galaxies of 
the whole system color-coded according to the weighting parameter 
$\delta_{V,i}/\sigma_V$, where $\delta_{V,i}$ is the difference
between the local mean velocity (computed on a group formed by
the $i$-th galaxy and the 10 closest neighbors) and the global mean velocity 
(see also Section~\ref{memb}). The positions of the two BCGs are marked with two black
crosses. For visual reference, we plot a circle of radius equal to $r_{200}=1.82$~Mpc.}
\label{figpvel}
\end{figure}

The presence of the Sext structure is also shown by our analysis of
the combined 3D information of the projected positions and line-of-sight
velocity. We used a modified version of the \citet{dre88} test, where the
mean velocity is considered separately from the velocity dispersion
(hereafter DS$\left<V\right>$-test). This test considers the
deviations of the local mean velocities from the global mean velocity, $\delta_{V,i}$, 
where the deviation is computed on the group formed by the $i$-th galaxy and 
its 10 closest neighbors \citep[e.g.][]{gir97,biv02,fer03,gir10,gir14}.
The DS$\left<V\right>$-test
detects significant substructures, at $>99.9\%$ c.l. using 1000
Monte Carlo simulated clusters. Figure~\ref{figpvel} shows in a color-coded plot the
resulting distribution of local mean-velocity deviations ($\delta_{V,i}$).
The Sext structure clearly stands out because of its
low velocity and it appears elongated toward the main system, while the main 
cluster is elongated along the NE-SW direction.

\begin{figure}
\centering 
\includegraphics[width=8.4cm]{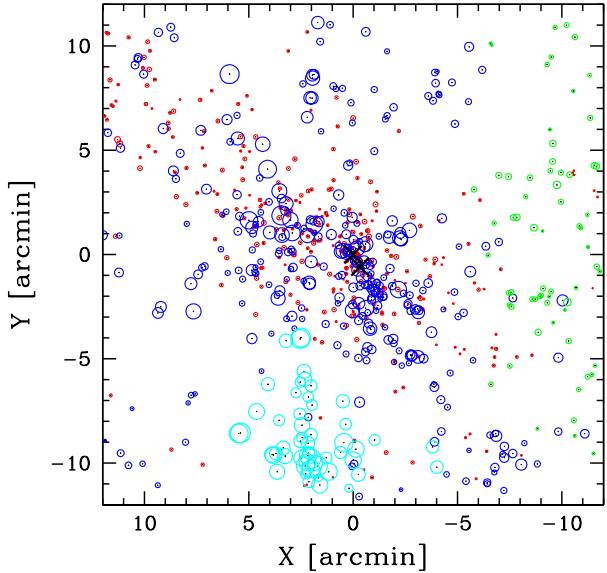}
\caption{Spatial distribution of the 781 cluster-member galaxies of 
the whole system. The size of the circles are weighted by the line-of-sight velocity of 
each galaxy.The positions of the two BCGs are marked with two black crosses. 
The four different colors indicate galaxies belonging to the 
four substructures detected through our 3D analysis, as described in the text: 
Lo-V (blue), Hi-V (red), W (green), and Sext (cyan).}
\label{3dstr}
\end{figure}

We applied the 3D-DEDICA method \citep{pis93,pis96} to the full spectroscopic 
sample. This method leads to a very complex description of the cluster structure, 
detecting 18 groups that are statistically significant at $\gtrsim99.99\%$ c.l., 
having at least ten galaxies, and with a relative density larger than 0.2, plus 
a plethora of less significant groups.
In order to refine the identification of galaxies of the Sext structure and, 
consequently, to obtain a ``cleaner characterization'' of the main system, we 
ran an alternative 3D analysis, using a simplified version of the 
3D-DEDICA method. Our simplified version is based on the same definition of the 
adaptive kernel estimate of \citep{pis93,pis96}, in particular with the same
computation of the local bandwidth factor $\lambda_i$ \citep[see Eqs.~26 and 27 in][]{pis93}, 
but where the size parameter $\sigma$ is not estimated using the recursive procedure of 
\citet[][see their method to obtain $\sigma_n$]{pis93,pis96}. We rather adopted the
rule-of-thumb value, $\sigma \sim 2.6 \sigma_n$, proposed by \citet{sil86}. We stress 
that our simplified procedure, while still using an adoptive kernel method,
is optimized to trace mainly the large-scale structure of the cluster.
The simplified procedure is successful in recovering the Sext structure 
(53 galaxies), as well as two main subclusters at low (Lo-V, 316 galaxies) and high
velocity (Hi-V, 331 galaxies) with $\Delta V_{\rm rf}\sim 1200$ \kss, plus a minor, less 
concentrated western structure (W, 81 galaxies), whose relative
density is only $\sim 10\%$, here reported for the sake of completeness.   
Table~\ref{tabdedica3d} lists the properties of the four substructures detected 
by our 3D analysis. The spatial distribution of these substructures is shown in 
Figure~\ref{3dstr} (see also Figure~\ref{figvd}, where the same colors are used for the 
four substructures).

\begin{table}
\caption{Results on the detection of 3D substructure in the whole sample
of cluster members.}
\label{tabdedica3d}
\begin{center}
\begin{tabular}{l r c c c r}
\hline
\hline
Sub-clump  & N$_{ S}$ &$V$ & $\alpha$(J2000) $\delta$(J2000) & $\rho_S$ & $\chi^2_S$ \\
 (1)      &  (2)     & (3)                             & (4)      & (5)  & (6) \\
\hline
Hi-V   & 331  &119907& 04:16:10.2 -24:04:04  & 0.98 &494 \\
Lo-V   & 316  &118201& 04:16:05.4 -24:05:32  & 1.00 &387 \\
W      &  81  &119136& 04:15:32.9 -24:05:39  & 0.13 & 89 \\
Sext   &  53  &116900& 04:16:18.8 -24:12:47  & 0.23 &105 \\
\hline
\end{tabular}
\end{center}
\begin{footnotesize} \textbf{Notes.} Columns list the following information: 
(1) id of sub-clump, (2) number of assigned members, (3) peak velocity in 
km~s$^{-1}$, (4) coordinates, (5) density relative to the densest peak, and 
(6) $\chi^2$ value for each peak.
\end{footnotesize}
\end{table}

We assume that the main system (MS) is formed by the two main subclusters 
plus the smaller W structure. The two main subclusters have similar galaxy content 
($\sim300$ galaxies) and roughly similar velocity dispersion $\sigma_V\sim500-550$ \kss.
Given the cluster mass computed within $r_{200}$=1.82~Mpc (see Section~5), if 
simply scaling the virial mass by $\sigma_V^2$, from the measured $\sigma_V$ for the two 
subclusters we obtain masses of $\sim 3-4\times 10^{14}$~M$_\odot$ within the virial region. The Sext 
structure lies at $\Delta z\sim -0.008$ from the MS, corresponding to a rest-frame 
velocity of $\Delta V_{\rm rf} \sim -1700$ \kss, and it is a low-mass structure, as 
indicated by the velocity dispersion of its galaxies \citep[$\sigma_{V,{\rm Sext}}\sim
300$ \kss, corresponding to a virial mass of $\sim 2\times 10^{13}$~M$_\odot$, following][]{mun13}.
The MS is clearly elongated along the NE-SW direction. We used the moments of inertia 
method \citep{car80,pli02} to compute ellipticity ($\epsilon$) and orientation ($PA$). 
We obtain $\epsilon=0.31\pm0.03$ and $PA=63\degr\pm3\degr$ (measured
counter-clock-wise starting from the north).

When looking at the spatial distribution of the two main subclusters on the plane of 
the sky (see Figure~\ref{figk2g}, lower panel), we find that galaxies belonging to the high-velocity 
subcluster (Hi-V) clearly concentrate around the central overdensity C, while those 
belonging to the low-velocity subcluster (Lo-V) are separated into two clumps centered on the 
secondary SW peak and the central overdensity C, being separated by $\sim2'$ which 
corresponds to $\sim600$ kpc and with a density ratio of 1:0.7. Thus, the central 
spatial overdensity C contains both low- and high-velocity galaxies. 

The secondary SW peak is spatially coincident with a possible, secondary mass 
concentration detected in the combined strong and weak gravitational lensing map by 
\citet{jau15}, although we caution that, after careful re-analysis of combined strong- 
and weak-lensing mass reconstruction using the \citet{mer09} model, we do not detect any 
lensing signal coincident with the SW galaxy overdensity (see Section~\ref{dis}
and Appendix).
Given the absence of X-ray emission and their lensing detection, \citet{jau15} 
concluded that this secondary SW clump is likely a
non-virialized structure, possibly associated to a large-scale structure filament. However, 
the large number of spectroscopic redshifts has allowed us to find that the SW peak is, 
indeed, populated by a high concentration of low-velocity galaxies, comparable in number to 
the high-velocity galaxies around the C peak.
Indeed, the analysis of deeper {\em Chandra} data by \citet{ogr15} reveals the presence of 
X-ray emission out to the SW peak, where a discontinuity is detected in the density 
of the intra-cluster medium (ICM). Finally, we find no galaxy 
concentration around the secondary NE mass concentration reported by \citet{jau15}.

\begin{figure}
\centering 
\includegraphics[width=8.4cm]{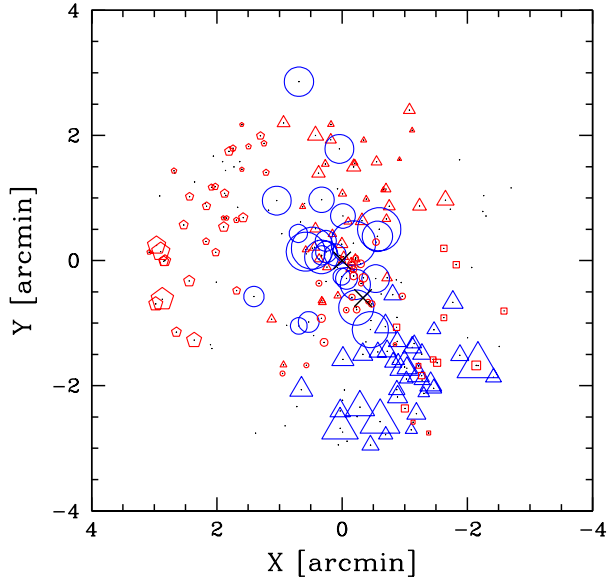}
\caption{Same as Figure~\ref{3dstr}, but showing the results of the full 3D-DEDICA 
procedure applied to cluster member galaxies within 1 Mpc from the cluster center. 
Different symbols refer to the six groups detected through the full 3D-DEDICA analysis, 
as listed in Table~\ref{1mpcded}: 1 (red triangles), 2 (blue triangles), 3 (red 
pentagons), 4 (blue circles), 5 (red circles), and 6 (red squares). Blue and red 
colors are assigned to groups having peak velocities close to the Lo-V and Hi-V 
subclusters, respectively. Blue circles and triangles highlight the complex 2D structure 
of the Lo-V subcluster.}
\label{1mpc}
\end{figure}

We also used the full 3D-DEDICA procedure to analyze the cluster center, within a 
radius of 1 Mpc, a region encircling 223 cluster galaxies. The results of this
analysis are shown in Figure~\ref{1mpc} and Table~\ref{1mpcded}. We detect six groups 
significant at the $\gtrsim 99.99\%$ c.l., having at least ten galaxies, and with a 
relative density larger than 0.15. All groups have peak velocities that are very close 
to those of the Lo-V or Hi-V subclusters, supporting the fact that the structure of 
the cluster can be interpreted, as a first approximation, as a bimodal one. 
The peak velocities of the four groups related to the Hi-V subcluster lie in a small 
velocity range ($\Delta V_{\rm rf}<300$ \kss). 
The two groups related to the Lo-V subcluster (group 2 and 4 in Figure~\ref{1mpc} and 
Table~\ref{1mpcded}) also have peak velocities that are not significantly separated 
according to both the 1D-DEDICA and 1D-KMM analysis. However, when analyzing the combined 
sample of 61 galaxies (36 plus 25 galaxies of group 2 and 4, respectively), the 
2D-DEDICA method indicates a clear bimodality with two, highly significant groups. This 
can be clearly appreciated also when looking at the distribution of blue circles and 
triangles in Figure~\ref{1mpc}. 
All our analyses restricted in the central regions provide additional support in 
favor of the complex structure of the Lo-V subcluster.

\begin{table}
\caption{3D-DEDICA results on substructures within 1 Mpc from the cluster center.}
\label{1mpcded}
\begin{center}
\begin{tabular}{l r c c c r}
\hline
\hline
Sub-clump  & N$_{ S}$ & $V$ & $\alpha$(J2000) $\delta$(J2000) & $\rho_S$ & $\chi^2_S$ \\
 (1)       &  (2)     & (3)          & (4)                    & (5)& (6)   \\
\hline
1 Red Tria. & 42 & 119725 &04:16:08.9 -24:04:04& 0.97 &23\\
2 Blue Tria.& 36 & 118189 &04:16:05.0 -24:05:33& 1.00 &17\\
3 Red Pent. & 33 & 120014 &04:16:18.4 -24:03:07& 0.30 &13\\
4 Blue Circ.& 25 & 118205 &04:16:09.9 -24:03:55& 0.44 & 8\\
5 Red Circ. & 20 & 120159 &04:16:08.4 -24:04:11& 0.92 &20\\
6 Red Squa. & 13 & 119954 &04:16:03.4 -24:05:39& 0.15 & 6\\
\hline
\end{tabular}
\end{center}
\begin{footnotesize} \textbf{Notes.} Columns list the following information: 
(1) id of sub-clump, (2) number of assigned members, (3) peak velocity in km~s$^{-1}$, (4)
coordinates, (5) density relative to the densest peak, (6) $\chi^2$ value 
for each peak. 
\end{footnotesize}
\end{table}

According to the indicator test by \citet{geb91}, the NE-BCG can be readily associated 
in velocity space to the Hi-V subcluster and its projected position is coincident with 
the C peak spatial overdensity. The SW-BCG, however, has a $>99\%$ 
peculiar velocity with respect to the velocity distribution of galaxies of the Lo-V 
subcluster and it is spatially separated from both the SW and C overdensities 
(see Figure~\ref{figk2g}, lower panel). 
Interestingly, the SW-BCG seems to be unrelated to any galaxy concentration, although it 
is coincident with the position of the secondary peak of X-ray surface brightness 
\citep[see][]{ogr15} and the SW mass halo detected via strong lensing analysis by 
\citet{gri15}.

To conclude, MACS0416 is far more complex than a bimodal merging system with two spatially 
and dynamically separated galaxy clumps in 3D, each traced by a BCG. The observational 
picture points toward an atypical dynamical configuration of the cluster core 
composed of two main substructures, one of which shows a displacement of both its BCG and the 
hot, X-ray emitting gas with respect to the position of galaxies dynamically bound to the 
same substructure. We will discuss in Section~\ref{dis} the possible scenarios that may 
lead to the observed spatial and dynamical configuration.

\section{Dynamical analysis: mass profiles} \label{s:mass}
%%%%%%%%%%%%%%%%%%%%%%%%%%%%%%%%%%%%%%%%%%%%%%%%%%%

In this section we derive the mass profile of MACS0416
from the dynamical analysis of the spectroscopic sample of cluster members 
described in Section~\ref{memb}. We also provide a comparison between the
mass profile derived from our dynamical analysis and those derived from strong
lensing \citep{gri15}, weak lensing \citep{ume14}, and a
combination of both \citep{ume15}, as well as those derived from X-ray data
(see Appendix; Ettori et al. in prep.).

As for the dynamical analysis presented in this paper, we constrain
the mass profile using two different and complementary methods: the
Modeling Anisotropy and Mass Profiles of Observed Spherical Systems
(MAMPOSSt) by \citet{mam13}, and the so called ``Caustic technique''
by \citet{dia97}.

The former method has been developed with the aim of breaking the
degeneracy within the Jeans equation between the
mass and the velocity anisotropy profiles. The code performs a
maximum likelihood fit of the distribution of galaxies in the
projected phase space in order to constrain the parameters that
describe the profiles. On the one hand, the method assumes a shape for the 3D velocity
distribution, dynamical equilibrium of the cluster, and it requires
parametrized models for the number density, the mass, and the velocity
anisotropy profiles. On the other hand, MAMPOSSt requires no binning,
differentiation, or extrapolation of the observables. Furthermore, MAMPOSSt does not
assume any shape for the distribution function in terms of energy and
angular momentum and it does not assume a Gaussian line-of-sight
velocity distribution. Finally, any parametrization for the mass
anisotropy and velocity profiles can be used with MAMPOSSt.

We consider the spectroscopic sample of cluster-member galaxies 
obtained as described in Section~\ref{memb}. We first check the 
incompleteness of our spectroscopic sample, since this could affect the results of our 
dynamical analysis. In order to maximize completeness 
we decided to restrict our analysis to the magnitude range $18\le\mathrm{Rc}\le 22.5$ 
and then to correct the sample of spectroscopic members for incompleteness as a function 
of magnitude. We also checked the completeness in radial bins from the cluster center and 
corrected for the spatial variation of completeness of our spectroscopic sample. However, 
we remark that, in the magnitude range selected, spatial completeness is approximately 
constant (less than $10-15\%$ variation radially) from the cluster center out to the 
largest radii probed.

\begin{figure}
\begin{center}
\includegraphics[width=9.0 cm]{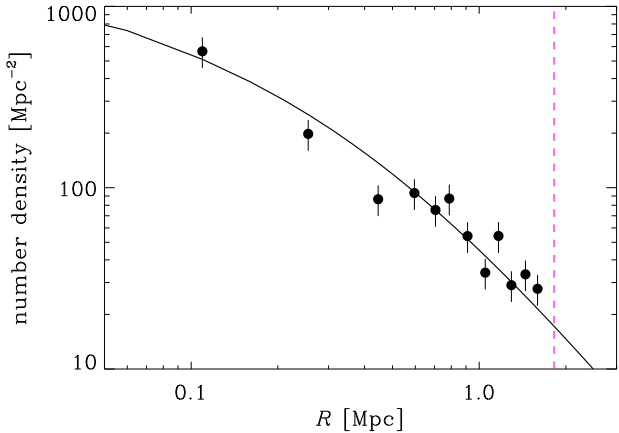}
\caption{Projected galaxy number density profiles $n(R)$ (black dots with
1$\sigma$ error bars) and best-fit projected NFW model (black solid line) for the whole
galaxy cluster spectroscopic members. The vertical dashed line indicates the value of
$r_{200}=1.82$~Mpc, obtained from the weak lensing analysis by \citet{ume14}.}
\label{fig:nd}
\end{center}
\end{figure}

We proceed by fitting the scale radius parameter of a projected NFW
\citep{nav97} density profile with a maximum likelihood technique. The
best-fit value for the scale radius of the galaxy deprojected density
profile is $r_{\nu} = 0.43\pm0.06$~Mpc.
In Figure~\ref{fig:nd} we show the projected number density profile for the
galaxy cluster spectroscopic members and the best-fit model. 

We use the MAMPOSSt technique to calculate the best-fit values
of the mass and anisotropy profile parameters. We restrict our analysis
to the virial region, where the cluster is more likely to have reached
dynamical equilibrium. For the virial region we use a region of radius 
$r_{200}= 1.82\pm0.11\,$Mpc, as estimated from the weak lensing by \citet{ume14}.

In our analysis we consider the \citet[NFW; ][]{nav97}, \citet{bur95}, \citet{her90}, and
\citet{ein65} mass profiles, that have been shown to provide good fits
to many cluster mass profiles \citep{bae02,rin06,biv06}, and
the Softened Isothermal Sphere \citep[SIS; ][]{gel99}. All these models,
except the \citet{ein65}, have two free parameters: the virial radius
$r_{200}$ and a scale radius $r_s$, which corresponds to the radius $r_{-2}$ for 
the NFW and \citet{ein65} models, where the logarithmic derivative of the mass density 
profile is $\rm{d\,ln}\rho/\rm{d\,ln}r=-2$. We fix the additional free parameter in 
the \citet{ein65} model to $m=5$, as in \citet{biv13}.

As for the velocity anisotropy profile, following \citet{biv13},
we consider three models:
\begin{itemize}
\item `C', constant anisotropy with radius, $\beta=\beta_C$;
\item `T': anisotropy profile from \citet{tir07};
\item `O': anisotropy of opposite sign at the center and at large radii.
\end{itemize}
The `C' model depends only on one parameter: the constant value of the velocity 
anisotropy $\beta_C$. The `T' and `O' models depend, instead, on two parameters: a 
scale radius $r_{\beta}$ and the anisotropy at a large radius, $\beta_{\infty}$. 
We fix $r_{\beta} \equiv r_{-2}$ in our models.

We consider only the `C' model in combination with the SIS mass
profile, because $r_{-2}$ cannot be uniquely defined for the SIS mass profile model. 
Therefore, we run MAMPOSSt on 13 combinations of
mass and velocity-anisotropy profile models, with three free
parameters: the virial radius $r_{200}$, the scale radius $r_s$ of the
mass profile, and $\beta_{\infty}$ or $\beta_C$. We use the
optimization routine NEWUOA \citep{pow04} to find the maximum
likelihood solutions.

The best-fit values of these parameters are presented in Table
\ref{tab:mam} along with their marginalized errors obtained 
by integrating the posterior probabilities over the other parameters.
We find that the best-fit model is obtained for a
combination of a SIS mass profile and a `C' velocity anisotropy
profile. Other combinations of models are statistically acceptable (at
the 1$\sigma$ level) according to the likelihood-ratio test by
\citet{mey75}. These are the models listed in Table
\ref{tab:mam}. The probability values of this test are listed in the
last column of Table \ref{tab:mam}. We also list the values
of the likelihood ratios relative to the best-fit model.  

\begin{table*}
\begin{center}
\caption{Results of the dynamical analysis with MAMPOSSt.}
\label{tab:mam}
\begin{tabular}{ccccccc}
\hline
\hline
Model              & $r_{200}$ & $r_s$                 & Vel.                   & $M_{200}$            & Like  & P($\chi^2$)\\
$M(r)$, $\beta(r)$ & [Mpc]     & [Mpc]                 & Anis.                  & [$10^{15}M_{\odot}$] & Ratio &     \\
(1)                & (2)       & (3)                   & (4)                    & (5)                  & (6)   & (7)  \\
\hline
SIS, C & $1.72_{-0.08}^{+0.09}$ & $0.02_{-0.01}^{+0.05}$ & $1.12_{-0.16}^{+0.46}$ & $0.88\pm0.13$ & $1.00$ & $-$ \\
Ein, C & $1.80_{-0.10}^{+0.08}$ & $0.68_{-0.20}^{+0.67}$ & $1.21_{-0.16}^{+0.51}$ & $1.01\pm0.14$ & $0.55$ & $0.25$ \\
NFW, C & $1.80_{-0.09}^{+0.08}$ & $0.63_{-0.19}^{+0.57}$ & $1.23_{-0.17}^{+0.52}$ & $1.01\pm0.14$ & $0.51$ & $0.28$ \\
Ein, T & $1.79_{-0.11}^{+0.08}$ & $0.58_{-0.13}^{+0.33}$ & $1.29_{-0.29}^{+1.56}$ & $0.99\pm0.15$ & $0.40$ & $0.39$ \\
NFW, T & $1.78_{-0.10}^{+0.08}$ & $0.53_{-0.13}^{+0.29}$ & $1.28_{-0.28}^{+1.52}$ & $0.98\pm0.15$ & $0.37$ & $0.42$ \\
Her, C & $1.83_{-0.09}^{+0.08}$ & $1.50_{-0.39}^{+1.07}$ & $1.28_{-0.17}^{+0.53}$ & $1.06\pm0.15$ & $0.36$ & $0.43$ \\
Ein, O & $1.85_{-0.11}^{+0.10}$ & $0.55_{-0.14}^{+0.38}$ & $0.91_{-0.14}^{+0.86}$ & $1.10\pm0.18$ & $0.30$ & $0.51$ \\
NFW, O & $1.83_{-0.10}^{+0.12}$ & $0.46_{-0.11}^{+0.30}$ & $0.98_{-0.21}^{+1.02}$ & $1.05\pm0.19$ & $0.27$ & $0.54$ \\
Bur, C & $1.83_{-0.07}^{+0.08}$ & $0.36_{-0.09}^{+0.20}$ & $1.31_{-0.19}^{+0.57}$ & $1.06\pm0.13$ & $0.21$ & $0.63$ \\
Her, T & $1.81_{-0.09}^{+0.07}$ & $1.23_{-0.24}^{+0.54}$ & $1.35_{-0.30}^{+1.70}$ & $1.03\pm0.14$ & $0.21$ & $0.63$ \\
Her, O & $1.89\pm0.11$          & $1.18_{-0.29}^{+0.68}$ & $0.84_{-0.11}^{+0.73}$ & $1.17\pm0.20$ & $0.15$ & $0.71$ \\
Bur, T & $1.80_{-0.08}^{+0.07}$ & $0.29_{-0.06}^{+0.12}$ & $1.29_{-0.27}^{+1.50}$ & $1.02\pm0.12$ & $0.11$ & $0.78$ \\
Bur, O & $1.83\pm0.08$          & $0.25_{-0.05}^{+0.07}$ & $0.97_{-0.19}^{+0.95}$ & $1.06\pm0.15$ & $0.07$ & $0.84$ \\
\hline 
\end{tabular}
\end{center}
\begin{footnotesize}\textbf{Notes.} Columns list the following information: (1)
model used for the mass profile, $M(r)$, and recipe used for the velocity anisotropies, 
$\beta(r)$, (2) $r_{200}$, (3) scale radius $r_s$, (4) derived velocity 
anisotropy, which is equal to $\beta_C$ for the 
C model and $\beta_{\infty}$ for the T and O models, (5) mass within $r_{200}$, (6) 
likelihood ratios relative to the best-fit model, and (7) $\chi^2$ probability, as 
obtained using the likelihood-ratio test, that the considered model is a worse 
representation of the data compared to the best-fit model (SIS+C). The uncertainties 
listed in this table refer to the 1$\sigma$ marginalized errors on each of the free 
parameters in our MAMPOSSt analysis. The scale radius is $r_{-2}$ for the NFW and Einasto 
models, $r_H=2r_{-2},r_B \simeq 2/3r_{-2},$ and the core radius, for the Hernquist, 
Burkert, and SIS $M(r)$ models, respectively.
\end{footnotesize}
\end{table*}

\begin{figure}
\begin{center}
\includegraphics[width=8.6 cm]{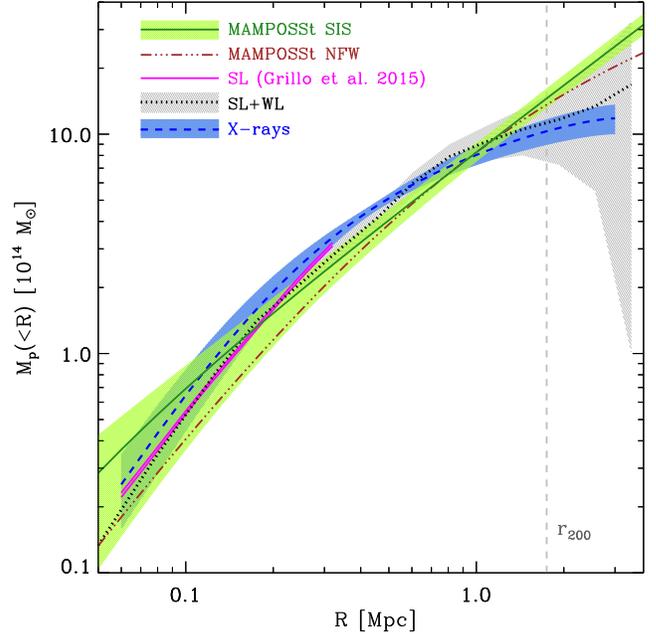}
\caption{Projected cumulative mass profile $M_{\rm p}(<R)$ with 1$\sigma$ confidence 
intervals for the MAMPOSSt projected SIS solution (green solid line and hatched region) 
and for the projected NFW solution (brown dash-triple-dotted line), both derived from 
a simple spherical component. For comparison, we 
plot the high-precision strong-lensing mass reconstruction by \citet{gri15} (magenta solid  
lines) and its combination with the weak lensing reconstruction by \citet{ume14} 
(black dotted line and gray hatched region), as well as the mass profile from our X-ray 
analysis (blue dashed line and hatched region). The dashed vertical line indicates
the value of $r_{200}$ obtained from our dynamical analysis with the SIS model. The
position of the NE BCG (R.A.=04:16:09.14, Dec.=-24:04:03.1) has been adopted as the 
center of the cluster for all the different profile reconstructions plotted here.}
\label{fig:comb}
\end{center}
\end{figure}

The best-fit mass profile model for MACS0416 is the SIS model. This model has been shown
to provide unacceptable fits to the mass profiles of clusters from the CAIRNS 
\citep[Cluster and Infall Region Nearby Survey; ][]{rin03} and the ENACS 
\citep[ESO Nearby Abell Cluster Survey; ][]{kat04} surveys. Most probably the mass profile of 
MACS0416 is different from the average cluster mass profile, because it is clearly
not a dynamically relaxed cluster. We remind that MAMPOSSt assumes dynamical relaxation, 
which might not be appropriate in the case of MACS0416. However, as shown in 
Figure~\ref{fig:comb}, despite the dynamically complex and clearly unrelaxed status of
the cluster, especially in the core region, the best-fit mass profile found by MAMPOSSt 
is in relatively good agreement with the mass profiles obtained from all the other 
independent probes and mostly within the 1$\sigma$ uncertainties. More specifically, 
in Figure~\ref{fig:comb} we plot for comparison the mass profile reconstruction from 
the strong lensing (SL) analysis by \citet{gri15}, a combination of the weak lensing 
(WL) reconstruction by \citet{ume14} and the SL model by \citet{gri15}, and the mass 
profile reconstructed from the X-ray data. The X-ray mass profile has been obtained 
under the assumption of a spherically-symmetric intra-cluster medium in hydrostatic 
equilibrium with the underlying gravitational potential, which might also not be 
appropriate for an unrelaxed cluster. A brief description of the method used to obtain 
the X-ray mass profile can be found in the Appendix. We notice that the largest deviations 
are found between the dynamical profile and the X-ray profile around 
$\sim300-400$~kpc from the cluster center. These deviations are most probably due 
to the presence of the secondary peak of X-ray emission associated with the SW BCG, 
which was not removed in the measurement of the hydrostatic X-ray mass presented here.
Given the high X-ray temperature, it is not possible to mask this secondary peak 
without a loss of information. We are currently working on a reconstruction of the 
X-ray mass which includes a reliable modelization of this secondary X-ray halo. We also
remind that throughout our analysis the 
cluster center has been fixed on the NE BCG, which marks the location of the bottom of 
the potential well of the system as shown by the strong lensing mass reconstruction 
\citep{gri15}. This is true for all the different probes of
the mass profiles shown in Figure~\ref{fig:comb}. In particular, the mass 
reconstructions from SL and WL are slightly different from those previously published 
because of the slightly different choice of the center.

The relatively good agreement among mass profiles obtained from different probes,
especially around the virial radius, leads to a good match between estimates of 
the cluster virial mass from different probes: for instance, when comparing the 
deprojected values the best-fit SIS model of MAMPOSSt yields a virial mass of 
$M_{200,c}= 8.8\pm1.3\times 10^{14}$~M$_\odot$, 
%$M_{200,c}= 15.2_{-1.5}^{+1.7}\times 10^{14}$~M$_\odot$, 
which is consistent within the uncertainties with the value obtained through the 
present SL+WL analysis, $M_{200,c} = 11.2\pm2.6 \times 10^{14}$~M$_\odot$ 
\citep[$M_{200,c} = 10.7\pm2.6 \times 10^{14}$~M$_\odot$ in the SL+WL analysis presented 
in][]{ume15}.
%$M_{200,c} = 11.4\pm4.0 \times 10^{14}$~M$_\odot$.
  
In order to extend the determination of the cluster mass profile beyond the virial 
region, we use the Caustic technique. This technique aims to identify density 
discontinuities in the distribution of galaxies in the plane of rest-frame, line-of-sight 
velocity vs. projected radii. Discontinuities are identified using, not only cluster 
members, but all the galaxies of the sample. The Caustic technique offers the advantage of 
allowing to calculate the mass of a cluster without assuming any parametrization for the 
profile and without the assumption of dynamical equilibrium. Following
\citet{dia99}, one has to solve the following equation:
\begin{equation}
M(<r) - M(<r_{min}) = \frac{1}{G} \int_{r_{min}}^{r} A^2(x) F_{\beta} (x) dx
\end{equation}
 where $A$ is the amplitude of the Caustic curve that is directly related to the
cluster potential, $G$ is the gravitational constant
and $F_{\beta}$ is 
\begin{equation}
F_{\beta} (r) = -2 \pi G \frac{\rho (r) r^2}{\phi(r)} \frac{3-2 \beta(r)}{1- \beta (r)}
\end{equation}
where $\rho (r)$ is the cluster mass density profile and $\phi(r)$ the gravitational 
potential of the cluster.

We first determine the amplitude $A$. In Figure~\ref{figvd}, we show the isodensity 
curves in the rest-frame line-of-sight velocity vs. projected radius plane, computed 
using a Gaussian adaptive kernel with an ``optimal'' kernel size, following \citet{sil86}. 
Although theory defines the Caustic as the curve of infinite density, in practice we 
proceed following \citet{dia99}. We use the determination of $r_{200}$ from the lensing 
analysis of \citet{ume14}. We select all galaxies with a rest-frame velocity 
$-6000\le V_{\rm{rf}} \le 6000$ km~s$^{-1}$ and we mirror the data with respect to the y axis 
(at $R=0\,$Mpc). This last step is necessary to suppress any edge-effect. 
Then, we calculate the isodensity profile and, finally, we define the Caustic as the 
curve that, within $r_{200}$ minimizes $\mid
  <v(esc, R)^2> - 4 \sigma_v^2 \mid$, where $<v(esc, R)^2>$ is the mean
  square escape speed determined from the Caustic amplitude within R,
  and $\sigma_v$ is the velocity dispersion of the member galaxies.
%  contains galaxies with line-of-sight velocities that are 
%as close as possible to the velocity dispersion of members. 
Then, we symmetrize the 
Caustic with respect to the zero-velocity axis, by choosing, in each radial bin,
the smaller absolute value of the Caustics among the two, positive and negative velocities.  

Once we have obtained the amplitude of the Caustic, $A$, we calculate the normalization 
$F_{\beta}$. The weakness of the Caustic method is that it needs to assume a value of 
$F_{\beta}$, which generally comes from simulations, while its strength is that it can 
be used beyond the virial radius. Since the mass profile of MACS0416 within the virial 
radius has already been reconstructed using different methods, we use the Caustic method 
to probe the mass profile at larger cluster-centric distances. 

In particular, we can determine the value of $F_{\beta}$. This can be obtained by 
imposing that the Caustic-derived mass value between $R=0$ and $R=r_{200}$ equals the 
value obtained with MAMPOSSt \citep[this technique has been first described and applied 
to the cluster MACS~J1206.2-0847 in][]{biv13}. For MACS0416 we find $F_{\beta}= 0.55$, 
which is in good agreement with results of \citet{dia97} and \citet{dia99}, while most 
recent implementations of the caustic algorithm by \citet{ser11} and \citet{gif13} favor a 
slightly higher value ($F_{\beta}= 0.68$). On the other hand, a value as low as
0.5 has still been used recently by \citet{gel13}. It is possible
that there is a real cluster-to-cluster variance in the value of
$F_{\beta}$, probably related to an intrinsic variance in the shape
of the different cluster mass profiles.
Hydrodynamical simulations have shown that the scatter on $F_{\beta}$ is large 
\citep[see Figure 4 of ][]{ser11}. We also stress that, given the unrelaxed nature of 
MACS0416 and the assumptions of spherical symmetry and dynamical equilibrium in the 
MAMPOSSt mass profile reconstruction, our determination of $F_{\beta}$ may be affected by 
systematics in this particular cluster.

Once we have determined $A$ and $F_{\beta}$, 
we can use the Caustic method to extend the mass profile reconstruction beyond $r_{200}$. 
The uncertainties on this profile are obtained by convolving the error on the measure 
of $M_{200}$ obtained from MAMPOSSt, with the error on the Caustic $M(<r)$, as
described in \citet{dia99}, but multiplied by a factor 1.4 to
have $\sim 1 \sigma$ errors \citep[in fact, the original prescription by
\citet{dia99} lead to estimate 50\% confidence levels, see][]{ser11}.
%the error on the Caustic
%$M(<r)$, as described in \citet{dia99}, with the error on the measure of $M_{200}$ 
%obtained from MAMPOSSt.

\begin{figure}
\begin{center}
\includegraphics[width=8.8 cm]{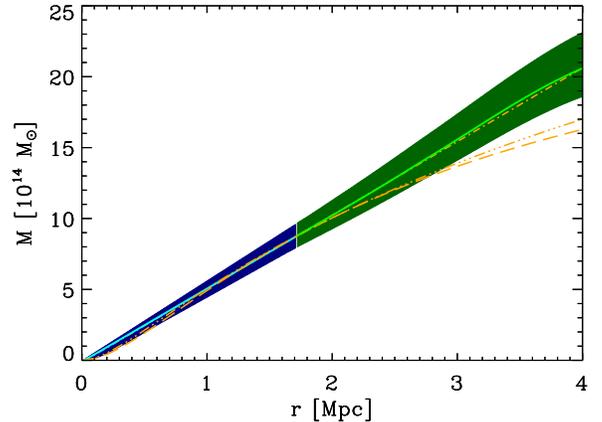}
\end{center}
\caption{Mass profile of MACS0416 cluster (with 1$\sigma$ uncertainties) as obtained 
from the dynamical analysis. The MAMPOSSt best-fit solution (SIS) and its 1$\sigma$ 
error are represented by the cyan solid curve and blue region. The MAMPOSSt best-fit 
solution is also shown extended beyond $r_{200}$ (yellow dash-dotted line). 
The Caustic profile and its 1$\sigma$ error are represented by the green solid line and 
the green region. For comparison we also show an NFW (three dotted-dashed orange line) 
and a Hernquist (long dashed orange line) profiles, with the same $r_{200}$ as the 
MAMPOSSt best-fit solution and $r_s$ as given in Table \ref{tab:mam}.}
\label{fig:dyn}
\end{figure}

The final mass profile obtained from the dynamical analysis, resulting from
the combination of the MAMPOSSt and Caustic techniques, is shown in Figure~\ref{fig:dyn}. 
In this plot we also compare our mass profile reconstruction with the NFW, Hernquist, 
and SIS models, all using the same value of $r_{vir}=1.77$ Mpc (the best-fit MAMPOSSt 
solution for the SIS model) and the values of $r_s$ listed in Table \ref{tab:mam}. 
This plot shows how, at large radii, the mass profile of MACS0416 is still well described 
by a SIS model, while it significantly deviates from the NFW or Hernquist model.

%__________________________________________________________________

\section{Discussion}\label{dis}

\begin{figure*}
\begin{center}
\plottwo{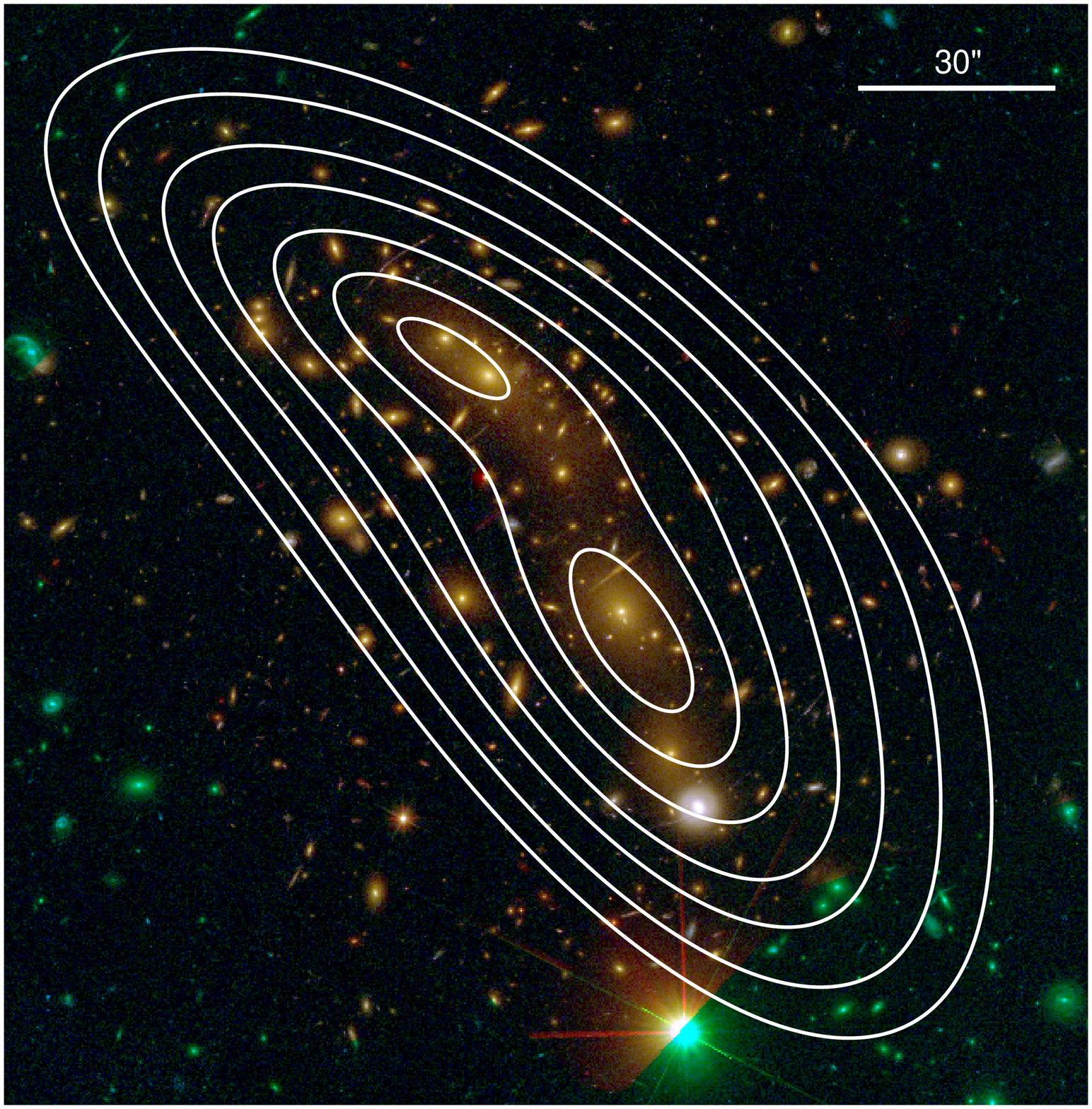}{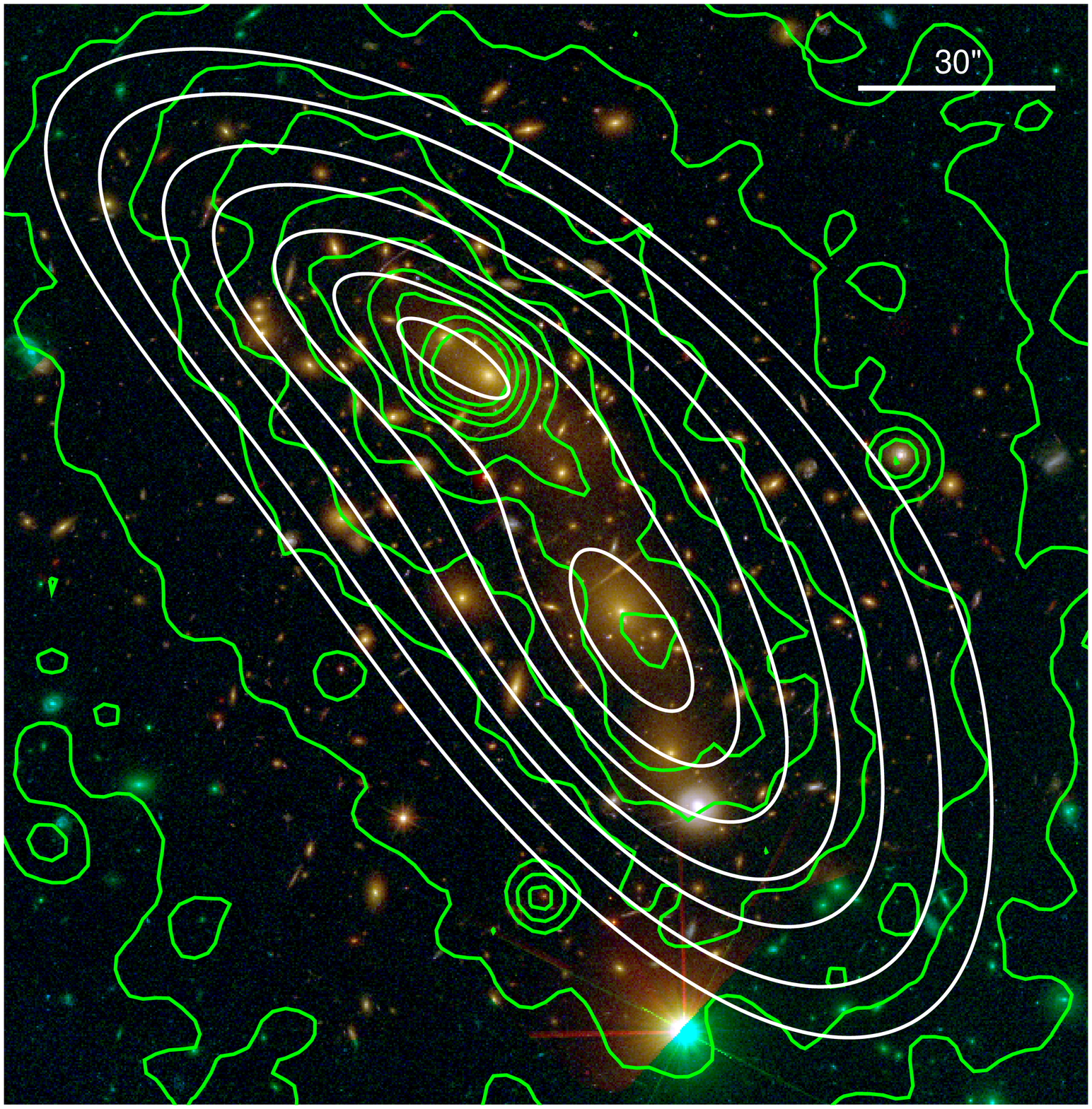}
\vspace{0.5cm}
\plottwo{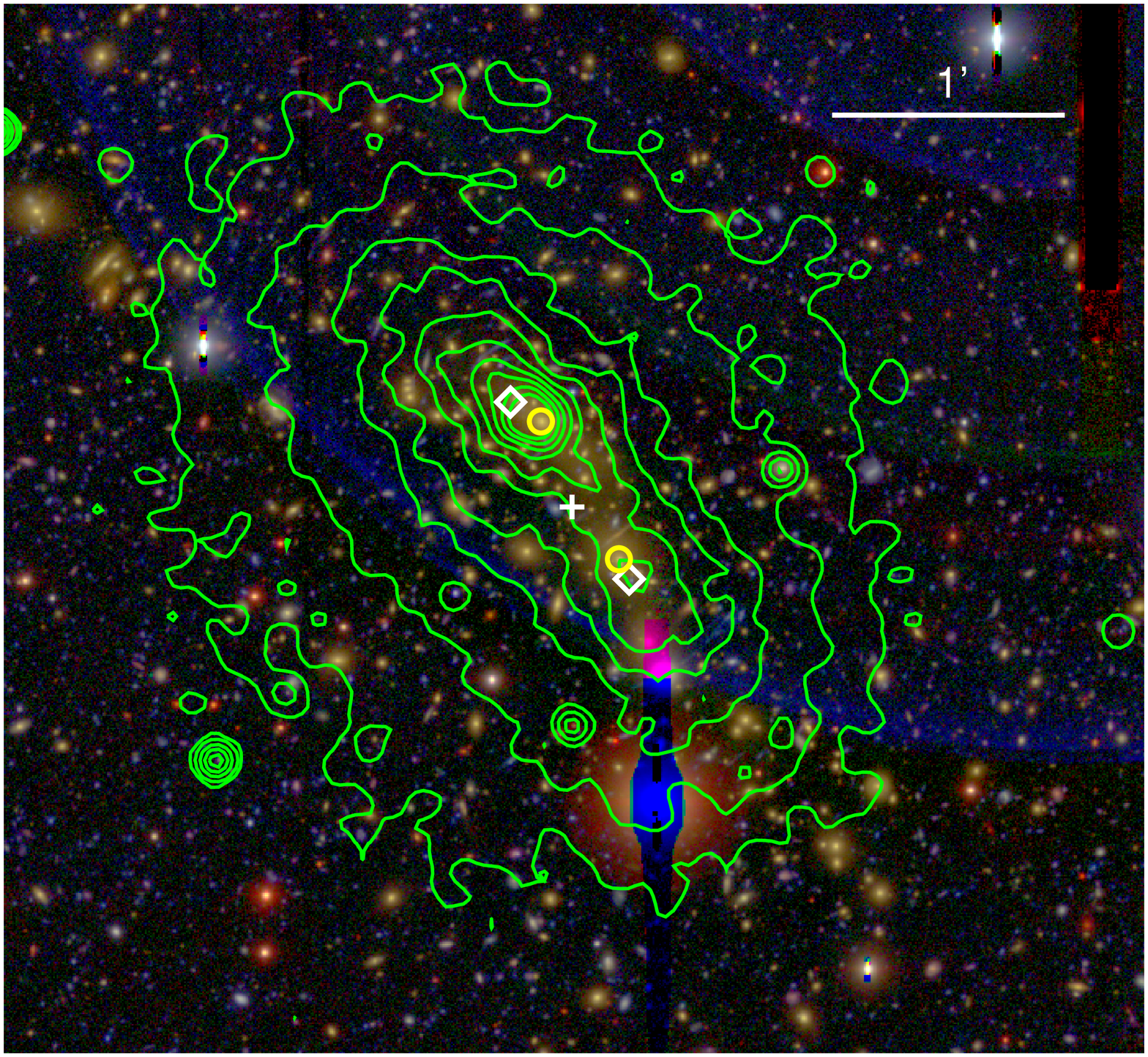}{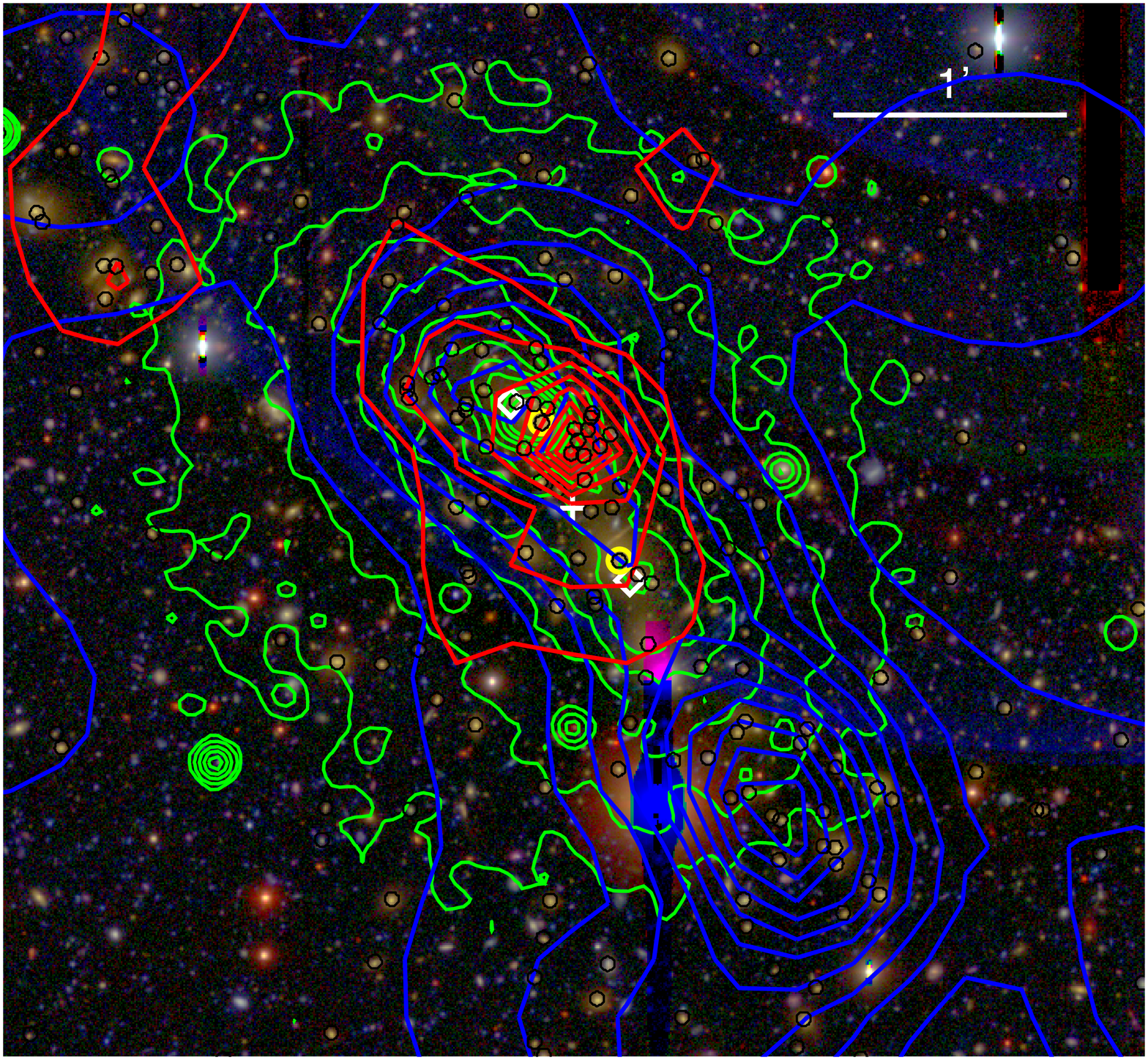}
\end{center}
\caption{\textit{Top left:} HST color image with extended mass halo from 
strong-lensing modelling \citep[white, ][]{gri15} contours overlaid.
\textit{Top right:} same as top left Figure, but with the addition of
$0.5-2$ keV X-ray \citep[green, ][]{ogr15} contours overlaid. 
\textit{Bottom left:} Subaru color image with X-ray contours overlaid (green). 
The positions of the two extended dark-matter halos (white diamonds) and the barycenter 
(white cross) calculated by \citet{gri15} are marked. The positions of the two BCGs are 
indicated with two yellow circles.
\textit{Bottom right:} same as in the bottom left Figure, but with the galaxy 
2D-isodensity contours of the two main subclusters from our 3D analysis overlaid. Blue 
and red isodensity contours are for galaxies of the low- (Lo-V) and high-velocity (Hi-V) 
subclusters, respectively. Black circles mark cluster member galaxies of our 
spectroscopic sample.}
\label{xraystr}
\end{figure*}

The offset between the collisional (hot gas) and non-collisional (stars and dark matter) 
mass components in merging galaxy clusters can be used to constrain the dark-matter 
cross section \citep[e.g.,][]{mar04,har15} as well as the dynamical configuration of the 
merger. A significant offset between the hot, X-ray emitting gas and dark matter, 
generally traced by lensing, is a clear signature of a post-merging scenario, while no 
offset between the two mass components suggests either a pre-merging phase or a 
configuration aligned along the line-of-sight. Recent results from 
deeper Chandra data of MACS0416 showed that both of the detected X-ray peaks 
(NE and SW) have very small displacement (a few arcsec in projection) with respect to 
the position of the two extended dark-matter halos detected through 
gravitational lensing \citep[see ][]{ogr15}. We 
verified that this is true also when adopting the more recent, high-precision strong 
lensing mass reconstruction by \citet{gri15}, as shown in Figure~\ref{xraystr}. Although 
the small offset seems to support a pre-merging scenario, it could still be 
consistent with a post-merger phase, if the merging of the two main subclusters has 
occurred along the line-of-sight.

From our dynamical and structural analysis we find that galaxies belonging to the 
NE and SW subclusters, as well as the two BCGs, are well separated in the projected 
velocity space (as clearly shown in Figure~\ref{fighistocl}), with a difference of 
projected line-of-sight velocities of the order of $900$~km~s$^{-1}$, where the NE 
subcluster has higher velocity than the SW one. This, together with the lack of a 
significant offset between dark matter and hot gas, could suggest that the merger is 
oriented along the line-of-sight. However, there are at least two additional 
complications in this scenario: 1) the low-velocity subcluster (Low-V) is split into 
two distinct, well separated ($\sim600$~kpc) sub clumps (C and SW) and 2) the SW-BCG is 
isolated, i.e. offset from any galaxy concentration. 

We now discuss possible interpretations able to reconcile the X-ray results 
\citep{ogr15} with those of our structural analysis in a consistent picture.

The X-ray temperature of the hot ICM of this massive cluster is high 
($\sim10$ keV), with no evidence of a cooling core and a very high value of the central
entropy \citep{don14}. This is expected in merging clusters, since the central hot gas has 
been shock-heated and has not had enough time to cool yet.

The sound speed in a cluster with a temperature of $\sim10$ keV is $\sim1300$~km~s$^{-1}$, 
while typical collision velocities are $1-2$ times the sound speed. The observed 
projected velocity difference between the two BCGs 
($\Delta V_{\rm{rf}}\sim900$~km~s$^{-1}$), as well as the projected velocity difference 
between the two main subclusters ($\Delta V_{\rm{rf}}\sim1200$~km~s$^{-1}$), together 
with the evidence that the merging is very likely happening at an intermediate angle 
between the l.o.s. and the plane of the sky, imply that the collisional velocity could 
well be close to, or above supersonic. 
The absence of a cool component in the X-ray gas around the cluster core (corresponding 
to the X-ray emissivity peak and the projected C galaxy overdensity) seems to suggest 
that the cluster core must have already experienced some perturbation. We also notice
that the X-ray surface brightness presents at least two density discontinuities in the SW
direction toward the secondary X-ray emission peak: the first one around the SW BCG and 
the second one toward the SW galaxy overdensity \citep[see ][]{ogr15}.

Together with the absence of a cool gas component, the presence of an X-ray cavity around 
the NE BCG, with no radio emission associated, provides an additional important clue. 
In a post-merging scenario, a cavity could have had enough time to form since first 
core passage. However, given the short timescale and the intense turbulence following 
the first core passage, forming a new cavity would have required a strong AGN outburst 
and the right timing. The expectation would be to observe radio emission associated 
with the X-ray cavity, while this is not observed. 
Instead, the presence of the X-ray cavity itself supports a pre-merger scenario. The 
cavity was most likely inflated by a recent weak outburst of the AGN in the NE BCG.

As shown by \citet{ogr15} the diffuse radio emission detected both in the JVLA and GMRT 
data extends on a size of $\sim0.6$~Mpc, that is smaller than that of the typical radio 
halos in merging clusters \citep[$1-1.5$~Mpc; e.g., ][]{fer12}. 
The size of the radio halo in MACS0416 is more similar to that of radio mini-halos 
typically found in cool-core clusters \citep[e.g., ][]{gia14}, but no evidence of a cool 
core is found from the X-rays. The radio halo seems to be associated to both subclusters.
It could be a single halo originating from the main merger event or it could be a 
superposition of two individual halos, where each of the two halos must have been 
generated by previous merger events within the two subclusters themselves.

Indeed, another possible scenario could be that of multiple mergers occurred over the past 
$\sim1$~Gyr, which is also postulated by numerical simulations \citep{poo08}. In this 
case, while the cluster core would have already relaxed back to a compact state, it 
should not have had enough time to cool back again. Some cool gas is expected to survive
the merger, but it could be difficult to detect given the different, unknown emission 
measure of the cold and hot ICM components, which in this case are probably overlapping 
in projection.

The dynamics of the galaxies belonging to the two main merging subclusters is complex.
The velocity and spatial distribution of galaxies is not the typical bimodal distribution 
observed in merging clusters. We find a clear spatial displacement between the SW BCG and
the bulk of the Lo-V subcluster to which it seems to be dynamically bound.
As our 3D analysis has shown, the Lo-V subcluster is highly substructured, being 
composed by two spatial overdensities of galaxies (separated by $\sim600$~kpc)
and an isolated BCG sitting halfway between the central overdensity and the SW sub-clump. 
A possible scenario able to explain the observed complex spatial and dynamical 
configuration, as well as the possible X-ray density discontinuity observed toward the 
SW sub-clump \citep[see][]{ogr15}, would be the following: the Lo-V subcluster could be 
itself the result of a previous or ongoing merger occurring at an angle close to the 
plane of the sky, while the two main subclusters (Lo-V and Hi-V) are most likely in a 
pre-merging phase. Although to a lesser extent, the NE part of the cluster could have 
been also affected by some merging, given the small concentration of galaxies detected 
at NE in the Hi-V subcluster (see Fig.~\ref{figk2g}). 

In order to assess the significance of the detection of the SW galaxy overdensity 
in the lensing shear signal, we analysed a lensing mass reconstruction obtained using a 
multi-scale grid lensing model that combines weak- and strong-lensing reconstruction 
following the methodology of \citet{mer09}. The surface-mass-density contours obtained 
with this model show no significant over-density around the relevant cluster members 
galaxies, indicating that the projected mass of the SW subclump must be significantly 
smaller than previously estimated by \citet{jau15}. The results obtained with this 
lensing analysis and some additional checks on their robustness are provided in the 
Appendix.

In conclusion, the complex structure we detect in the core of MACS0416 is consistent with 
a scenario in which the two main subclusters are in a pre-merging phase, although 
each subcluster could be the result of a recent merger, more dramatic in the Lo-V 
subcluster. Recent hydrodynamical modeling of the gas and dark matter by \citet{die15} 
supports our pre-merger conclusion, demonstrating that relatively small offsets between 
gas and dark matter are expected at the early stage of encounter due to gas compression 
when the separation of the two main components is less than the combined sum of their 
virial radii. The displacement of the SW-BCG could be explained by dynamical friction or 
by close encounter and tidal interaction with some other massive galaxy in the cluster 
core. An analogous example of a displaced, isolated BCG is that of the Coma cluster 
\citep[see ][]{biv96,neu03}.
Notice also that the ICL, clearly detected in both ground-based and HST images, is 
elongated approximately along the line connecting the two BCGs and closely following the 
elongation of the X-ray emission (see Fig.~\ref{xraystr}).

%Given the unrelaxed dynamical status of the cluster core and the observed deviation from 
%``universality'' of its mass profile, it may be interesting to investigate the possible systematic 
%which could introduce deviations from the theoretical predictions for the mass-concentration 
%relation (c-M). 
%We derived constraints on the mass and concentration parameters obtained when 
%fitting the mass profile with a NFW and constant $\beta$ model instead of the best-
%fitting model discussed in Section~\ref{s:mass}. Although the best fit value is 
%slightly offset toward smaller concentrations we find that, within the uncertainties, 
%the observed parameters are consistent with theoretical expectations from simulations.

%Figure~\ref{mc} shows the constraints on the mass and concentration parameters obtained when 
%fitting the mass profile with a NFW and constant $\beta$ model. Although the best fit value is 
%slightly offset toward smaller concentrations, within the uncertainties the observed parameters 
%are consistent with theoretical expectations from simulations.

%\begin{figure}
%\begin{center}
%\includegraphics[width=9.0 cm]{c_M.ps}
%\end{center}
%\caption{Confidence contours for the mass and concentration parameters obtained
%from fit with a NFW and a constant velocity anisotropy profile ($\beta$).
%The solid red line and shaded orange region represent the theoretical
%c-M relation of \citet{bha13} for all halos and its 1 sigma
%scatter. The dashed green line represents the theoretical
%c-M relation from \citet{deb13} for all halos.}
%\label{mc}
%\end{figure}

Given the unrelaxed dynamical state of the cluster core, it is worth to briefly 
discuss the results of our dynamical analysis. We have shown that for this complex 
merging cluster the mass profile is best described by a SIS model, rather than by a 
NFW profile. This may point toward interesting deviations from universality during 
major-merger phases in the cluster assembly history. However, one cluster is certainly 
not enough to derive solid conclusions on the effect of mergers on the total mass profile, 
as well as on the biases affecting different mass probes. We defer further investigation 
of the unrelaxed clusters of our CLASH-VLT sample and comparison with hydrodynamical 
simulations to a future work.

Further refinement of the dynamical analysis presented in this work, in particular 
abandoning a single spherical component approximation \textit{thus reducing systematics}, 
is underway and it will be presented in a future publication (Sartoris et al., in prep.). 

%__________________________________________________________________

\section{Conclusions}\label{con}

We performed a detailed dynamical and structural analysis of the Frontier Fields cluster
MACS0416 based on 781 spectroscopically confirmed cluster members. 

Our analysis shows that the cluster structure is more complex than that of a bimodal
merger. The most likely emerging scenario is the one where the two main subclusters 
are seen in a pre-merging phase, although each subcluster could be the result of a 
recent merger, which is more evident in the Lo-V subcluster. 
The displacement of the SW-BCG could be explained by dynamical friction or by close 
encounter and tidal interaction with some other massive galaxy in the cluster core. 

The cluster mass profile reconstructed through our dynamical analysis is best fit by a
Softened Isothermal Sphere model, significantly deviating from a NFW profile, 
especially at large radii (beyond the virial radius). Despite the complex and clearly 
unrelaxed nature of this massive cluster, the mass profile reconstruction obtained 
through our dynamical analysis is in good agreement with those obtained from strong and 
weak lensing, as well as with that from the X-rays. 

%A refined dynamical analysis using two subcomponents for the two main merging 
%subclusters aimed at reducing the systematics inherent in the analysis presented here, is
%currently underway (Sartoris et al., in prep.). 
Separate studies focused on the influence of the environment on different galaxy 
populations and their interplay with the substructures of this cluster will be 
presented in subsequent papers \citep[][Mercurio et al. in prep.]{mai16}.

\begin{acknowledgements}
We thank the anonymous referee for the careful reading of the manuscript and 
the valuable comments and suggestions provided.
We acknowledge the continuous help of the ESO user support group, especially 
our project support astronomer, Vincenzo Mainieri, for his excellent support and advices. 
We acknowledge financial support from PRIN-INAF 2014 1.05.01.94.02 and from MIUR 
PRIN2010-2011 (J91J12000450001). I.B. acknowledges funding 
support from the European Union Seventh Framework Programme 
(FP7/2007-2013) under grant agreement n$^\circ$~267251 ``Astronomy Fellowships in Italy'' 
(AstroFIt). B.S. acknowledges a grant from ``Consorzio per la Fisica - Trieste''.
P.R. acknowledges the hospitality and support of the visitor program of the DFG cluster 
of excellence ``Origin and Structure of the Universe''. G.A.O. and A.Z. are supported by 
NASA through Hubble Fellowship grants HST-HF2-51345.001-A and \#HST-HF2-51334.001-A, 
respectively, awarded by the Space Telescope Science Institute, which is operated by the 
Association of Universities for Research in Astronomy, Incorporated under NASA contract 
NAS5-26555. J.M. is supported by the People Programme (Marie Curie Actions) of the 
European Union Seventh Framework Programme (FP7/2007-2013) under REA grant agreement 
number 627288.
This work is based on data collected at the ESO VLT (prog.ID 186.A-0798), 
at the NASA HST, and at the NASJ Subaru telescope. Based in part on data collected 
at Subaru Telescope and obtained from the SMOKA, which is operated by the Astronomy 
Data Center, NAOJ \citep{bab02}.
\end{acknowledgements}

%__________________________________________________________________

\begin{appendix}
\section{Determination of the X-ray mass profile}

\begin{figure}
\begin{center}
\includegraphics[width=9.0 cm]{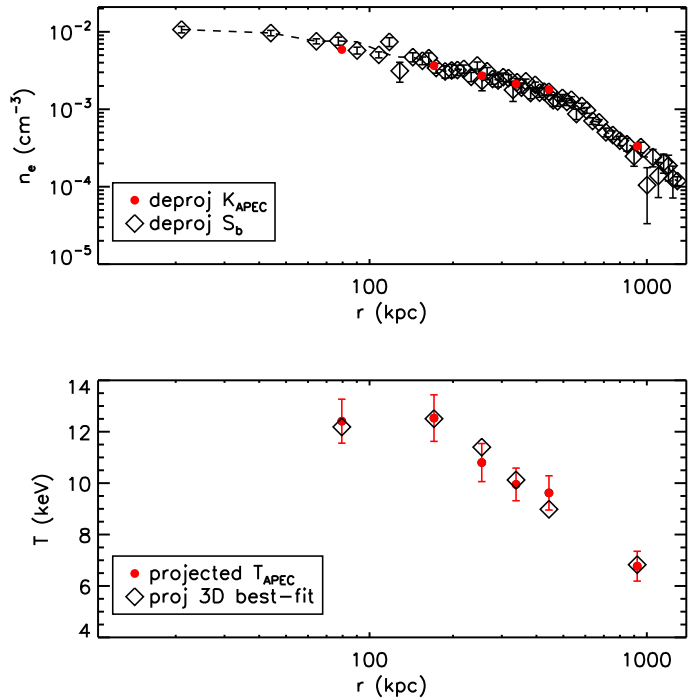}
\end{center}
\caption{\textit{Top panel:} electron density profiles obtained from both the 
geometrically-deprojected surface brightness profile (black diamonds) and the 
normalization of the thermal component fitted to the extracted spectra (red circles). 
\textit{Bottom panel:} best-fit spectral measurements of the gas temperature 
(red circles). The black diamonds represent the projection in the spectral bins of 
the best-fit 3D temperature profile obtained by inverting the hydrostatic equilibrium 
equation in combination with the deprojected surface brightness profile (black diamonds 
in the top panel).}
\label{Tnprof}
\end{figure}

The X-ray mass profile has been obtained under the assumption of a spherically-symmetric 
intra-cluster medium in hydrostatic equilibrium with the underlying gravitational 
potential. None of the substructures identified in the optical analysis has been 
masked. We treated the whole X-ray emission, including the secondary X-ray peak of 
surface brightness, as originating from hot gas in hydrostatic equilibrium in a single 
dark matter halo. Chandra observations (obsID: 
16236, 16237, 16304, 16523, 17313; total exposure time: 293 ksec) have been reduced and 
analyzed using CIAO 4.7 and CALDB 4.6.9 to recover a gas density profile from the 
deprojection of the cumulative surface brightness profile and a gas temperature profile 
with a local background. The top panel of Figure~\ref{Tnprof} shows the electron 
density profiles obtained from 
both the geometrically-deprojected surface brightness profile and the normalization of 
the thermal component \citep[APEC model in Xspec 12.9;][]{arn96} fitted to the extracted 
spectra. The gas temperature has been obtained by spectral fitting with an APEC 
model in Xspec 12.9. The bottom panel of Figure~\ref{Tnprof} shows the best-fit 3D 
temperature profile obtained by inverting the hydrostatic equilibrium 
equation in combination with the deprojected surface brightness profile.
A functional form \citep[the King approximation to the isothermal sphere in the case 
discussed here; see e.g.][]{ett02} for the gravitational potential has been assumed and 
constrained in its 2 free parameters (normalization and scale radius) by using the 
observed gas density profile and the hydrostatic equilibrium equation to match the 
spectroscopic temperature profile (i.e. we apply the backward method described in 
\citet{ett13}. In the mass profile reconstruction presented in Figure~\ref{fig:comb}, 
at each radius the errors on the mass profile represent the range of 
values allowed from the $1\sigma$ statistical uncertainties on the 2 free parameters 
(i.e. $\Delta \chi^2 =2.3$). More details on 
the method used here for the mass reconstruction are presented in \citet{ett10}. 

\section{Combined strong- and weak lensing mass reconstruction}

\begin{figure}
\begin{center}
\includegraphics[width=9.0 cm]{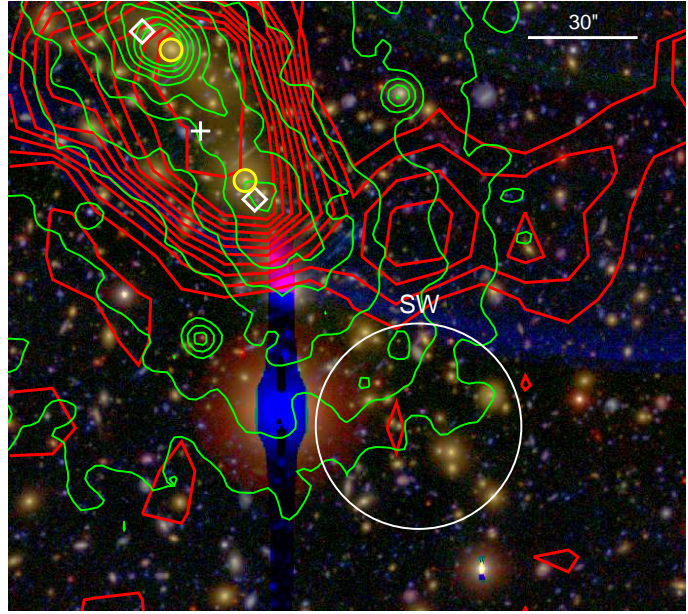}
\end{center}
\caption{Subaru color image with overlayed convergence contours (red) obtained 
from the combined strong- and weak-lensing model described in the text. X-ray contours 
are also displayed in green. The positions of the two extended dark-matter halos 
(white diamonds) and the barycenter (white cross) calculated by \citet{gri15} are marked. 
The positions of the two BCGs are indicated with two yellow circles. The position of the 
SW overdensity of cluster member galaxies is indicated with a white circle.}
\label{SWWL}
\end{figure}

We analysed a lensing mass reconstruction covering the full cluster field and in 
particular the area of interest around the SW substructure. We used a multi-scale grid 
lensing analysis that combines weak- and strong-lensing reconstruction following the 
methodology of \citet{mer09}. This is the same method used in the analysis of the X-ray 
relaxed CLASH sample presented in \citet{mer15}. This lens model of MACS0416 was 
submitted in the context of the pre-Frontier Fields lens models provided by 
STScI (http://www.stsci.edu/hst/campaigns/frontier-fields/Lensing-Models) and 
uses CLASH Subaru/Suprime-Cam weak-lensing catalogs \citep{ume14}, HST ACS weak-lensing 
catalogs \citep{mer15,zit15} and the CLASH strong-lensing identifications presented in 
\citet{zit15}. As shown in Figure~\ref{SWWL}, the surface-mass-density contours, 
which are derived from 1000 bootstrap realisations of this model, 
reveal no significant over-density around the position of the 
SW concentration of member galaxies. In order to test the robustness of this result 
which, in this area of the field, is mainly driven by the HST/ACS weak-lensing catalog, 
we performed two checks. First, we used our comprehensive spectroscopic redshift catalog 
to validate the weak-lensing background selection for the ACS catalog. We find no 
contamination by foreground objects in the area of interest, which confirms our initial 
selection based on the CLASH 16-band photometric redshifts. We also performed a 
reconstruction which is based on an ACS catalog which has no background selection applied 
at all, but was just cleaned from stars and artefacts. We do this to increase the number 
of galaxies in the weak-lensing analysis. Although this selection may introduce a 
dilution of the lensing signal, it ensures that no existing substructure is missed due 
to insufficient spatial resolution. However, also the reconstruction 
based on this extreme case scenario shows no significant overdensity in the reconstructed 
surface-mass density map.

\section{Redshifts of X-ray sources in the field}

\begin{figure}
\begin{center}
\includegraphics[width=9.0 cm]{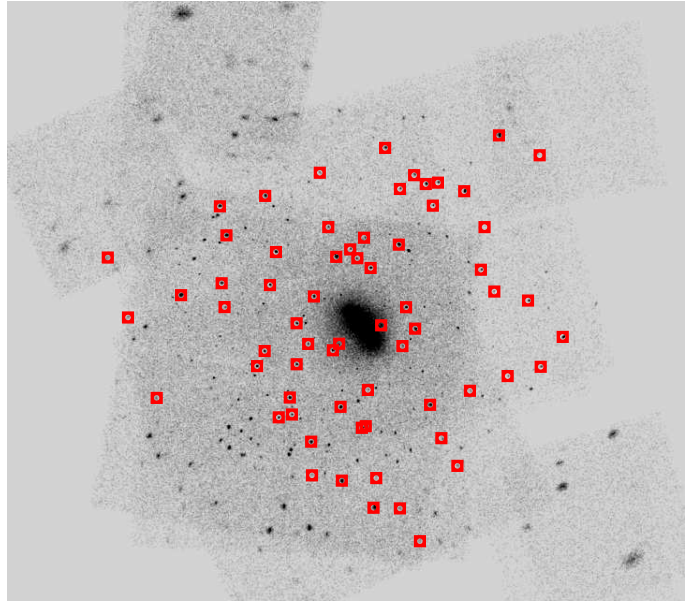}
\end{center}
\caption{Chandra $0.5-7$ keV image with overlaid X-ray sources and serendipitously 
discovered AGN with spectroscopic redshift from our CLASH-VLT survey. The Chandra image 
shown here is not exposure corrected and not background subtracted, to best show the 
detected point sources. Notice that only a handful of the sources in our spectroscopic 
catalog are cluster members, while most are foreground and background sources 
(see Table~\ref{xrayt}).}
\label{xray}
\end{figure}

In our CLASH-VLT survey we have systematically targeted all the Chandra X-ray sources 
detected in the field of each cluster. In the field of MACS0416, we have obtained 
redshifts and spectra of 60 Chandra sources, plus 5 serendipitously discovered AGN. 
Table~\ref{xrayt} lists coordinates, magnitudes, and redshifts of these objects. 
Figure~\ref{xray} shows the position of all the sources with spectroscopic redshift in 
the Chandra field.
 
\begin{table*}
\begin{center}
\caption{Redshifts of X-ray sources in the field of MACS0416.}
\label{xrayt}
\begin{tabular}{lcccccc}
\hline
\hline
ID & RA & DEC & $z$ & QF & Ref. & Mag \\
(1) & (2) & (3)    & (4)   & (5) & (6) & (7) \\
\hline
\hline 
\multicolumn{2}{l}{Chandra Sources}\\
 CLASHVLTJ041524.4-240449  & 63.851613 & -24.080402 & 0.3150 & 3 & 1 & 19.49 \\
 CLASHVLTJ041529.3-240620  & 63.871900 & -24.105615 & 0.6399 & 3 & 1 & 22.32 \\
 CLASHVLTJ041532.1-240259  & 63.883668 & -24.049812 & 2.0723 & 3 & 1 & 21.13 \\
 CLASHVLTJ041536.6-240648  & 63.902366 & -24.113490 & 0.3547 & 3 & 1 & 19.95 \\
 CLASHVLTJ041538.6-235438  & 63.910659 & -23.910696 & 2.3129 & 3 & 1 & 18.92 \\
 CLASHVLTJ041539.5-240232  & 63.914738 & -24.042268 & 0.3984 & 3 & 1 & 22.51 \\
 CLASHVLTJ041542.5-240126  & 63.927177 & -24.023894 & 0.4015 & 3 & 1 & 21.07 \\
 CLASHVLTJ041546.2-235727  & 63.942583 & -23.957776 & 1.6857 & 3 & 1 & 19.50 \\
 CLASHVLTJ041547.7-241121  & 63.948598 & -24.189198 & 0.3037 & 3 & 1 & 20.48 \\
 CLASHVLTJ041551.2-240957  & 63.963448 & -24.165885 & 0.5691 & 2 & 1 & 22.01 \\
 CLASHVLTJ041552.0-235702  & 63.966847 & -23.950565 & 1.4959 & 3 & 1 & 22.51 \\
 CLASHVLTJ041553.2-235811  & 63.971504 & -23.969948 & 1.9899 & 3 & 1 & 23.33 \\
 CLASHVLTJ041553.7-240815  & 63.973841 & -24.137775 & 2.4928 & 3 & 1 & 22.50 \\
 CLASHVLTJ041554.7-235706  & 63.978027 & -23.951751 & 0.3075 & 3 & 1 & 19.26 \\
 CLASHVLTJ041556.0-241509  & 63.983157 & -24.252706 & 0.4193 & 3 & 2 & 20.43 \\
 CLASHVLTJ041557.1-240425  & 63.987727 & -24.073615 & 0.0000 & 3 & 1 & 15.28 \\
 CLASHVLTJ041557.3-235639  & 63.988751 & -23.944318 & 2.9691 & 3 & 1 & 21.23 \\
 CLASHVLTJ041559.1-240320  & 63.996110 & -24.055575 & 0.3903 & 3 & 1 & 20.18 \\
 CLASHVLTJ041559.9-240517  & 63.999418 & -24.088297 & 0.5690 & 3 & 1 & 22.65 \\
 CLASHVLTJ041600.4-235721  & 64.001827 & -23.955929 & 1.0167 & 3 & 1 & 20.67 \\
 CLASHVLTJ041600.4-241330  & 64.001638 & -24.225091 & 0.3439 & 3 & 1 & 20.05 \\
 CLASHVLTJ041600.7-240010  & 64.002833 & -24.002840 & 1.4572 & 3 & 1 & 19.48 \\
 CLASHVLTJ041603.7-235516  & 64.015484 & -23.921314 & 1.7844 & 3 & 1 & 21.26 \\
 CLASHVLTJ041604.6-240414  & 64.019065 & -24.070821 & 0.4111 & 3 & 1 & 19.24 \\
 CLASHVLTJ041606.2-241328  & 64.026017 & -24.224534 & 0.5697 & 3 & 3 & 22.13 \\
 CLASHVLTJ041606.9-240120  & 64.028810 & -24.022490 & 2.1517 & 9 & 1 & 24.27 \\
 CLASHVLTJ041607.5-240730  & 64.031228 & -24.125260 & 0.3550 & 3 & 2 & 19.58 \\
 CLASHVLTJ041608.0-240920  & 64.033244 & -24.155826 & 2.1311 & 3 & 1 & 21.87 \\
 CLASHVLTJ041608.4-235949  & 64.034888 & -23.997007 & 0.3068 & 3 & 1 & 23.49 \\
 CLASHVLTJ041608.8-240925  & 64.036819 & -24.157175 & 2.1915 & 3 & 3 & 23.87 \\
 CLASHVLTJ041609.8-240051  & 64.041017 & -24.014281 & 0.5605 & 9 & 1 & 24.05 \\
 CLASHVLTJ041611.4-240024  & 64.047656 & -24.006897 & 0.1535 & 3 & 1 & 18.03 \\
 CLASHVLTJ041613.3-241206  & 64.055284 & -24.201708 & 1.3670 & 2 & 2 & 21.68 \\
 CLASHVLTJ041613.5-240822  & 64.056455 & -24.139513 & 2.0760 & 3 & 1 & 22.05 \\
 CLASHVLTJ041613.9-240510  & 64.057793 & -24.086357 & 0.4992 & 6 & 5 & 20.75 \\
 CLASHVLTJ041614.5-240047  & 64.060523 & -24.013144 & 0.3520 & 3 & 3 & 20.21 \\
 CLASHVLTJ041615.3-240530  & 64.063601 & -24.091894 & 2.2072 & 3 & 1 & 23.74 \\
 CLASHVLTJ041616.3-235917  & 64.067802 & -23.988083 & 0.4627 & 2 & 1 & 22.17 \\
 CLASHVLTJ041618.2-235632  & 64.075664 & -23.942274 & 0.3906 & 3 & 1 & 21.04 \\
 CLASHVLTJ041619.5-240247  & 64.081083 & -24.046522 & 0.3955 & 3 & 1 & 19.43 \\
 CLASHVLTJ041620.1-241007  & 64.083577 & -24.168858 & 1.3647 & 3 & 1 & 23.16 \\
 CLASHVLTJ041620.7-240511  & 64.086382 & -24.086521 & 0.3991 & 5 & 4 & 20.09 \\
 CLASHVLTJ041623.3-240408  & 64.097065 & -24.069136 & 2.1201 & 3 & 1 & 22.90 \\
 CLASHVLTJ041623.3-240613  & 64.096987 & -24.103652 & 2.4718 & 3 & 1 & 22.96 \\
 CLASHVLTJ041624.3-240845  & 64.101395 & -24.145976 & 0.2677 & 3 & 3 & 20.33 \\
 CLASHVLTJ041624.8-240753  & 64.103322 & -24.131653 & 1.6901 & 3 & 1 & 19.65 \\
 CLASHVLTJ041627.2-240854  & 64.113505 & -24.148384 & 0.4661 & 3 & 1 & 20.30 \\
 CLASHVLTJ041627.9-240032  & 64.116082 & -24.009063 & 1.9569 & 3 & 1 & 20.21 \\
 CLASHVLTJ041629.2-240213  & 64.121604 & -24.036946 & 4.1494 & 3 & 1 & 19.69 \\
 CLASHVLTJ041630.2-235742  & 64.125989 & -23.961766 & 1.1927 & 3 & 1 & 19.67 \\
 CLASHVLTJ041630.4-240532  & 64.126515 & -24.092477 & 0.3380 & 3 & 1 & 20.28 \\
 CLASHVLTJ041632.0-240618  & 64.133494 & -24.105221 & 0.7110 & 3 & 1 & 22.48 \\
 CLASHVLTJ041638.8-235942  & 64.161665 & -23.995040 & 2.3779 & 3 & 1 & 23.47 \\
 CLASHVLTJ041639.2-240319  & 64.163437 & -24.055474 & 0.6332 & 3 & 2 & 24.52 \\
 CLASHVLTJ041639.9-240207  & 64.166129 & -24.035398 & 2.1216 & 3 & 1 & 21.15 \\
 CLASHVLTJ041640.2-235814  & 64.167684 & -23.970560 & 0.6217 & 3 & 1 & 22.32 \\
 CLASHVLTJ041648.8-240243  & 64.203525 & -24.045421 & 1.1785 & 3 & 1 & 21.93 \\
 CLASHVLTJ041654.3-240754  & 64.226124 & -24.131936 & 0.5882 & 3 & 2 & 21.94 \\
 CLASHVLTJ041700.6-240350  & 64.252481 & -24.064163 & 0.3009 & 3 & 1 & 19.65 \\
 CLASHVLTJ041705.1-240049  & 64.271075 & -24.013635 & 0.5531 & 3 & 1 & 19.70 \\
\hline
\multicolumn{2}{l}{Serendipitous AGN}  \\
 CLASHVLTJ041529.6-235538  & 63.873166 & -23.927416 & 3.6528 & 3 & 1 & 21.84 \\
 CLASHVLTJ041541.7-235916  & 63.923836 & -23.988006 & 0.3559 & 3 & 1 & 22.46 \\
 CLASHVLTJ041544.9-240733  & 63.937133 & -24.125920 & 3.0600 & 3 & 1 & 22.94 \\
 CLASHVLTJ041605.6-241158  & 64.023505 & -24.199521 & 0.4497 & 3 & 1 & 22.65 \\
 CLASHVLTJ041619.9-241150  & 64.082823 & -24.197260 & 2.4718 & 3 & 1 & 24.10 \\
\hline
\end{tabular}
\end{center}
\begin{footnotesize}\textbf{Notes.} Columns list the following information: (1)
VIMOS identification number, (2-3) coordinates, , (4) spectroscopic redshift, (5) 
redshift quality flag, (6) reference (i.e. CLASH-VLT VIMOS: 
1=based on LR-Blue spectra, 2=based on MR spectra, 3=based on a combination 
of LR-Blue and MR spectra, 4=\citet{ebe14}, 5=Magellan, (Kelson, privat comm.)), and 
(7) Subaru R-band magnitude.
\end{footnotesize}
\end{table*}

\section{Redshifts of radio sources in the field}

We cross-matched our CLASH-VLT spectroscopic catalog with the catalog of JVLA detected
sources in the field of MACS0416. Allowing for a position uncertainty of $3''$, we 
found redshifts and spectra of 105 JVLA sources. Table~\ref{jvla} lists coordinates, 
magnitudes, and redshifts of these objects.

\begin{table*}
\caption{Redshifts of radio sources in the field of MACS0416.}
\label{jvla}
\begin{center}
\begin{tabular}{l c c c c c c c}
\hline
\hline
ID      & RA    & Dec & $z$ & QF & Ref. & Mag  \\
(1)     &  (2)    & (3)  & (4)   & (5) & (6) & (7)  \\
\hline
\hline
 CLASHVLTJ041705.4-241006  & 64.272324 &  -24.168458 & 0.2700 & 3 & 1 & 18.55 \\
 CLASHVLTJ041703.8-241121  & 64.266004 &  -24.189252 & 0.3027 & 3 & 1 & 18.94 \\
 CLASHVLTJ041703.5-240603  & 64.264658 &  -24.101109 & 0.5659 & 3 & 1 & 21.48 \\
 CLASHVLTJ041653.0-235435  & 64.220664 &  -23.909770 & 0.3983 & 3 & 1 & 19.63 \\
 CLASHVLTJ041652.0-235850  & 64.216827 &  -23.980742 & 0.4005 & 2 & 1 & 22.95 \\
 CLASHVLTJ041651.7-240546  & 64.215239 &  -24.096121 & 0.2052 & 3 & 1 & 19.89 \\
 CLASHVLTJ041648.8-240243  & 64.203525 &  -24.045421 & 1.1785 & 3 & 1 & 21.93 \\
 CLASHVLTJ041648.4-240034  & 64.201492 &  -24.009517 & 0.4125 & 3 & 1 & 20.21 \\
 CLASHVLTJ041647.7-235706  & 64.198886 &  -23.951909 & 0.3835 & 3 & 1 & 19.74 \\
 CLASHVLTJ041647.0-235307  & 64.195996 &  -23.885462 & 0.2994 & 3 & 2 & 19.87 \\
\hline
\end{tabular}
\end{center}
\begin{footnotesize}\textbf{Notes.} Only a portion of the table is shown here to 
illustrate its form and content. The entire table is available in the 
electronic edition of this Journal, on CDS, and at following URL: 
\textit{https://sites.google.com/site/vltclashpublic/}. The full table contains 7 
columns and 105 redshifts. 
Columns list the following information: (1) VIMOS identification number, (2-3) 
coordinates, , (4) spectroscopic redshift, (5) redshift quality flag, (6) reference 
(i.e. CLASH-VLT VIMOS: 
1=based on LR-Blue spectra, 2=based on MR spectra, 3=based on a combination 
of LR-Blue and MR spectra, 4=\citet{ebe14}, 5=Magellan (Kelson, private comm.)), and (7) 
Subaru R-band magnitude.
\end{footnotesize}
\end{table*}

\end{appendix}

\end{document}